\documentclass[11pt]{article}
\usepackage[margin=1in]{geometry}
\usepackage{setspace}
\usepackage{graphicx}
\usepackage{booktabs}
\usepackage{longtable}
\usepackage{array}
\usepackage{subcaption}
\usepackage{float}
\usepackage{amsmath, amssymb}
\usepackage{caption}
\usepackage{hyperref}
\usepackage{siunitx}
\usepackage{xcolor}
\hypersetup{colorlinks=true,linkcolor=purple,citecolor=magenta,urlcolor=blue}
\usepackage{multirow}
\usepackage{lscape}
\usepackage{threeparttable}
\usepackage{adjustbox}
\usepackage[backend=biber,style=apa,natbib=true]{biblatex}
\DeclareLanguageMapping{american}{american-apa} % optional
\usepackage{tabularx,makecell,array} % preamble
\newcommand{\sym}[1]{\ifmmode^{#1}\else\(^{#1}\)\fi}

\usepackage{tikz}
\newcommand{\tikzxmark}{%
\tikz[scale=0.23] {
    \draw[line width=0.7,line cap=round] (0,0) to [bend left=6] (1,1);
    \draw[line width=0.7,line cap=round] (0.2,0.95) to [bend right=3] (0.8,0.05);
}}

    \makeatletter
\def\@fnsymbol#1{\ensuremath{\ifcase#1\or \dagger\or \ddagger\or
   \mathsection\or \mathparagraph\or \|\or **\or \dagger\dagger
   \or \ddagger\ddagger \else\@ctrerr\fi}}
    \makeatother

\title{Who gets hit first and who recovers last? Evidence from Indian Coastal Flood Shock}
\author{Jheelum Sarkar\thanks{Department of Economics, American University. Email: js8622a@american.edu
}}

\date{\today}

\begin{document}
\maketitle
\begin{abstract}
Catastrophic floods directly risk 1.8 billion lives worldwide, most of whom are from East and South Asia. How do extreme floods reshape paid labor outcomes? To answer this, I focus on a 1-in-100 year flood event in India. I first combine Sentinel-1 SAR with JRC Global Surface Water dataset to generate flood map. Using information from this map in various rounds of periodic labor force surveys, I estimate gender-specific dynamic effects of the flood shock. Key results show that men experienced short-lived reduction in their employment while women faced a delayed but persistent decline in their working hours. Men suffered most in secondary sector and increased their participation in primary sector. Women were hit hardest in the tertiary sector. Such sectoral impacts could be attributable to disruptions in infrastructure and physical capital. Moreover, marital status and dependency burden further shape the gender differential effects of the extreme flood event. Results remain robust under alternative treatment definitions.

\end{abstract}

\newpage
\section{Introduction}
Global average temperature is set to increase by at least 1.5~°C in the next two decades (\cite{IPCC2021-AR6-WGI}). This, in turn, is likely to exacerbate the frequency and intensity of extreme weather events such as floods, cyclones, droughts, and heatwaves (\cite{IPCC2021-AR6-WGI}). In particular, flooding is most widespread among all weather-related disasters. 23\% of the world's population are exposed to 1-in-100 year flooding \footnote{Note that 1-in-100 year flood refers to an extreme flood event which has 1\% probability of occurrence in a year.} (\cite{Rentschler2022FloodExposurePoverty}). Nearly one in five of these people live in India (\cite{Rentschler2022FloodExposurePoverty}).  

In low- and middle-income countries, gender inequality is a major challenge which are worsened by exogenous shocks including extreme weather. Existing studies have shown that extreme events such as droughts and floods increase discriminatory behavior against women such as early marriages (\cite{CornoHildebrandtVoena2020}), intimate partner violence (\cite{AguilarGomezSalazarDiaz2025}), food insecurity (\cite{HadleyEtAl2023}; \cite{AsfawMaggio2017}), and migration (\cite{AfridiMahajanSangwan2022}). However, less attention has been paid to how such weather shocks, particularly, extreme floods, impact men’s and women’s participation in the labor market.

Extreme weather adversely affect labor demand by changing wages and productivity. In Philippines, typhoons reduced hourly wages of formal workers in affected municipalities (\cite{FranklinLabonne2019}). By disrupting agricultural productivity, droughts declined on-farm employment in Ethiopia (\cite{MusunguKubikQaim2024}). In Nepal, an extreme flood event reduced monetary returns to agricultural labor by 9-10\% (\cite{KamblePaudelMishra2024}). Adverse weather shocks also impact labor supply. In India, absenteeism increases in manufacturing sectors if both current and lagged temperatures increase beyond 35°C (\cite{SomanathanEtAl2021}). Even in developed countries, climate extremes reduce labor supply. In the United States, for example, flood risk reduced labor supply by 0.33\% during 1988-2018 (\cite{JiaMaXie2022_NBER30250}). In response to droughts, women faced 19\% reduction in their working days compared to men when drought hit in rural India (\cite{AfridiMahajanSangwan2022}). 

I estimate gender-specific dynamic effects of coastal flood on labor market outcomes. This not only captures the magnitude of flood impact over time but also points out who is affected first and for whom return to paid work takes longer time. Key results show that men's paid labor declined immediately when the flood hit but rebounded soon after. On the contrary, women faced delayed decline in their working hours and recovered slowly. These gender differences in onset and recovery from flood shock are driven by sectors of employment and gendered social norms. Women were worst impacted in tertiary or service sector while men faced most setback in secondary secondary sector. Gendered social norms are captured by marital status and household dependency ratio. Married women experienced reduction in their paid labor while unmarried women did not see any change. Married men, on the other hand, increased their participation soon after the flood. Similar trends are observed among men and women coming from households with high dependency: Women reduced their paid labor during post flood periods while men re-enter the employment after flood. Results remain consistent after increasing threshold to define highly exposed districts to flood. Moreover, the findings are robust after dropping border districts to account for spatial spillover effects.

In developing countries, most of the studies on this topic focused on droughts and extreme temperatures (\cite{AfridiMahajanSangwan2022};\cite{SomanathanEtAl2021}). Among few studies that looked at floods, the focus has largely been on intrahousehold re-allocation of time. After the 2017 Bangladesh floods, for instance, women increased time spent on income-generating activities while men spent more time on domestic responsibilities (\cite{VitellozziGiannelli2024}). This does not reveal what is happening in the men's and women's paid labor market during the post-flood period. In contrast, I focus on how paid labor outcomes change for men and women after flood. I bring the dynamic effects to the forefront which traces who is hit first and who is last to recover. This information is important because it can improve on both timing and targeting of disaster responses. 

Several limitations remain. First, flood-induced migration can be a key factor behind gender difference in the onset and recovery phase but I am unable to capture migration aspect due to lack of data availability. Second, absence of granular geographical location in the data source beyond district makes it difficult to define finer treatment group. Third, while the event study supports parallel trends in the main specifications, some pre-trends did exist either for men's or women's working hours when employment outcomes are divided into primary, secondary and tertiary sectors. Thus, caution should be taken while interpreting estimates in case of sectoral resilience. Nevertheless, the estimated coefficients are helpful in identifying trends in employment of men and women after flood in various sectors.

\textbf{Section Outline:} Section \ref{con} outlines the context which discusses the case study. Section \ref{data} describes data and treatment definition. Section \ref{ident} presents the identification strategy. Section \ref{results} discusses the results, followed by the conclusion in Section \ref{conclude}. 

\section{Context}\label{con}
Kerala is a southwestern coastal state in India and spans a coastline of 580 km. Due to its geographical positioning, it is highly vulnerable to multiple natural disasters such as floods, landslides, cyclones, and coastal erosion (\cite{PDNA2018Kerala}). The map of Kerala is shown in figure~\ref{fig:kerala}.

Kerala experienced a 1-in-100 year flooding in August 2018, the worst since 1924 \parencite{CWC2018Kerala}. It received the heaviest downpours between August 8-10, 2018, and August 14-19, 2018 \parencite{KeralaSDMA2018}. Apart from excessive rainfall, poor dam management worsened the flood-induced damage across the state \parencite{KeralaSDMA2018}. The floods resulted in 498 confirmed casualties and displaced over 1.4 million people, who sought refuge in relief camps (\cite{PDNA2018Kerala}). 

\begin{figure}[h]
    \centering
    \includegraphics[width=0.8\linewidth]{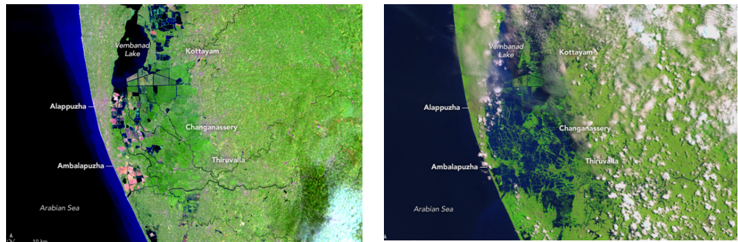}
    \caption{\textit{Pre- and Post-2018 Kerala Floods. The left panel shows Kerala on February 6, 2018, before the floods, while the right panel depicts the region on August 22, 2018, after the floods. Based on false-color imagery, here, bright green indicates vegetation; dark blue represents floodwaters visible in the post-flood image; darker hues in the pre-flood image denote permanent water bodies such as rivers and lakes; white color indicates cloud cover. \href{https://earthobservatory.nasa.gov/images/92669/before-and-after-the-kerala-floods}{NASA Earth Observatory images}.}}
    \label{fig:fig3}
\end{figure} 

The floods caused extensive damage to water supply infrastructure, disrupting urban water supply and contaminating drinking water sources. The estimated damage was initially valued at INR 19,512 crore (USD 2.8 billion) but was later revised to INR 24,308.35 crore (USD 3.5 billion) \parencite{KeralaSDMA2018}. According to satellite imagery, approximately 65,188 hectares of land were submerged (\cite{PDNA2018Kerala}).
\begin{table}[htbp]
\centering
\caption{Summary of Infrastructure Damages Due to the 2018 Kerala Floods}
\label{tab:kerala_damages}
\begin{threeparttable}
\begin{tabular}{p{7cm} p{5cm}}
\toprule
\textbf{Sector} & \textbf{Damages (Million USD)} \\
\midrule
Agriculture            & 2,159 \\
Livestock              & 24.6  \\
Fisheries              & 20.16 \\
Small-scale industries & 91.7  \\
Tourism                & 315.7 \\
\bottomrule
\end{tabular}
\begin{tablenotes}
\footnotesize
\item \textit{Source:} Self-compilation from the joint report on the 2018 Kerala Floods by \cite{PDNA2018Kerala}.
\end{tablenotes}
\end{threeparttable}
\end{table}

\newpage
\section{Data and Treatment Definition}\label{data}
 
\subsection{Periodic Labor Force Survey (PLFS):}
This paper utilizes quarterly data on paid labor outcomes from two rounds of Periodic Labor Force Surveys (PLFS) conducted during July 2017-June 2018 and July 2018-June 2019. PLFS is a nationally representative survey conducted by the National Statistical Office (NSO) of India to assess key labor market indicators, including employment, unemployment, and labor force participation. These surveys include demographic and socioeconomic information for all household members interviewed.

\textbf{Limitation:} Migration is an important channel which can affect the extent to which a climate event can impact labor market outcomes. However, PLFS did not include migration related information before 2022.
\subsection{Sentinel-1 SAR data and JRC Global Surface Water dataset}
Data on 2018 flood in Kerala were obtained from Copernicus Sentinel-1 satellite imagery using C-band synthetic aperture radar (SAR) backscatter on Google Earth Engine (GEE). To characterize flooded areas, I compared pre-flood (March 1-March 31, 2018) and flood-period (August 1-September 1, 2018) surfaces. Permanent water bodies were excluded using JRC Global Surface Water dataset. The resultant flood mask was aggregated at subdistrict level to generate flooded areas. Figure~\ref{fig:kerala_flood} gives districts of Kerala on the basis of their flood exposure. Details can be found in Appendix \ref{subsec:inundation}. 
\begin{figure}
    \centering
    \includegraphics[width=0.5\linewidth]{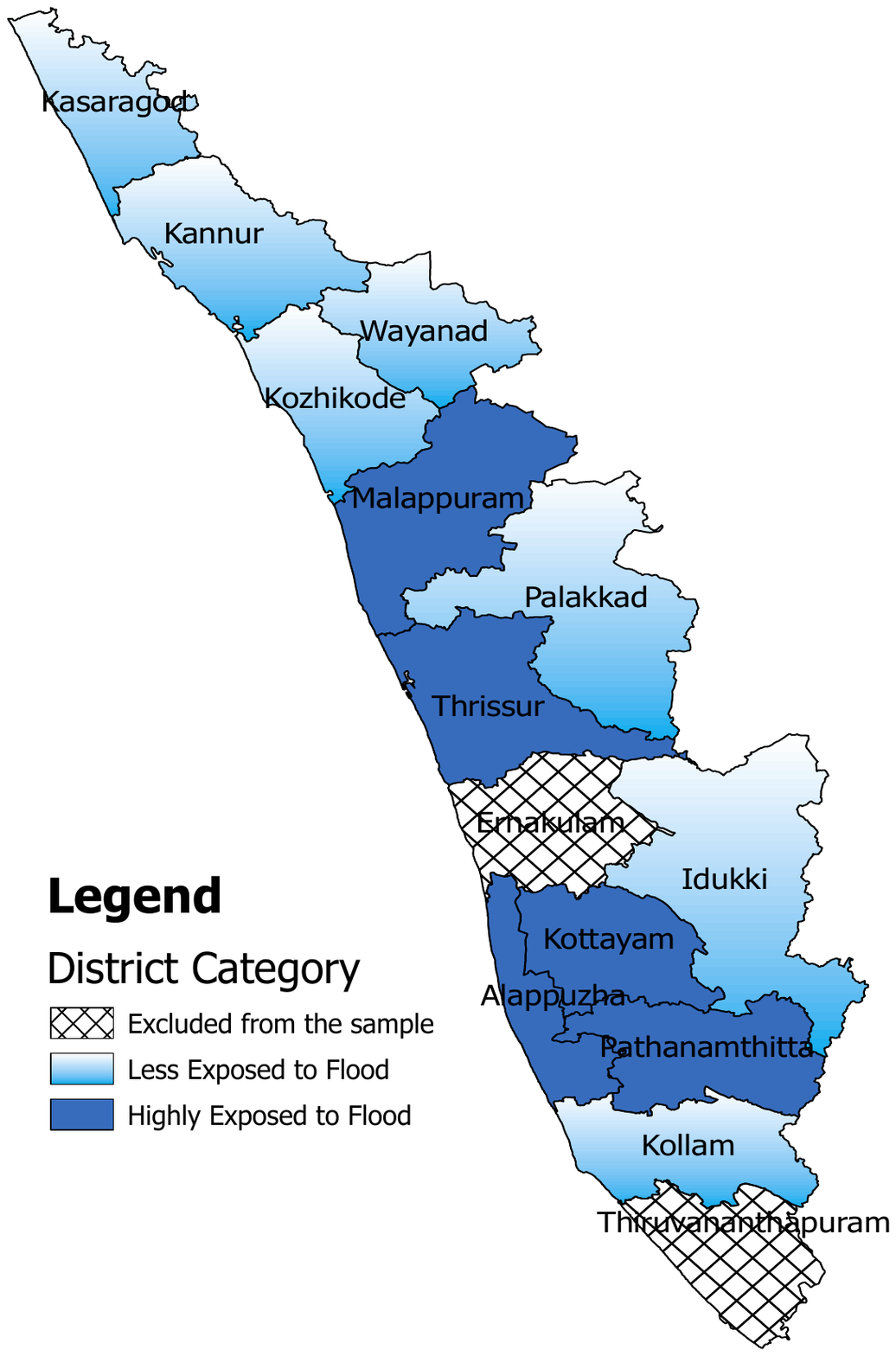}
    \caption{\textit{Kerala districts by flood exposure. Dark blue colored regions represent districts which are highly exposed to flood. Light blue colored regions denote districts which are less exposed to flood. Hatched districts are excluded from analysis.
    Source: Self-compiled using shapefiles from \href{https://github.com/geohacker/kerala/tree/master/shapefiles}{GitHub repository}.}}
    \label{fig:kerala_flood}
\end{figure}
\subsection{2011 Census data}
I use 2011 census to extract data on geographical areas of each subdistrict and district. Because PLFS data gives household location at district level, I aggregate subdistrict level inundation data for getting district level aggregates. Simply summing raw subdistrict level inundation data for each district, however, may not fully capture the flood exposure because larger inundated areas may simply be due to larger total geographic area. Hence, I used weighted summation method, where I define weight as share of each subdistrict level area within a district:
\begin{equation}\label{eqa}
\text{Exposure}_{d}
= \sum_{i \in \mathcal{I}_d}
\left( \frac{\text{Subdistrict Area}_{i}}{\text{District Area}_{d}} \times \text{Inundation}_{i} \right)
\end{equation}
\noindent\text{where} $\mathcal{I}_d$ denotes the set of subdistricts in district $d$.
This definition gives district-level exposure measure which uses geographical area based weights. That is, higher weights are given to larger subdistricts and lower weights to smaller subdistricts within each district. The exposure values are reported in table~\ref{tab:district_exposure} from Appendix. Because all districts were more or less affected by the flood, I used first Jenks natural breakpoint as threshold to classify districts as highly exposed and less exposed to the 2018 flood\footnote{Jenks natural breaks optimization minimize the sum of within-class squared deviations. With $\text{K}$=4 and flood exposure values for 12 districts, I get three optimal cut-points: $\text{91.15, 187.14, 464.85}$. This separates less exposed districts($\text{Exposure} < 91.15$) from relatively higher exposed districts ($\text{Exposure} \geq 91.15$)}.

\textbf{Treatment and Control Districts:}
Since less exposed districts may have had little prior exposure, this catastrophic flood could have substantial effects in such districts. Hence, less exposed districts may not be a clean comparison group. To address this, I include all districts of the neighboring state, Tamil Nadu as control. This choice is justified due to geographical proximity and comparable socioeconomic indicators, as indicated by low standardized mean differences in table~\ref{tab:sumstats_cohensd_pretreatment}. Thus, Tamil Nadu is a valid counterfactual, provided that the parallel trend holds. This is captured using event study results in \ref{results}. I utilize highly exposed and less exposed districts as alternative treatment groups separately to compare whether labor market outcomes vary by flood intensity in Kerala against the same control group.\\
It should be noted that a cyclone named ‘Ockhi’ passed by Kerala and Tamil Nadu on November 29, 2017 while heading towards Lakshadweep (\cite{RajyaSabha2018-Ockhi-PSC211}). This storm was short-lived and only confined to fishing communities of certain coastal areas. It impacted two districts (out of 14) in Kerala and four (out of 32 districts) districts in Tamil Nadu(\cite{RajyaSabha2018-Ockhi-PSC211} ; \cite{NHRC2018-Ockhi-TN}). To ensure a cleaner analysis, I excluded these districts which leaves me with 12 Kerala districts and 28 Tamil Nadu districts in the final sample.

\textbf{Why not a continuous flood exposure?}\\
Continuous treatment requires to satisfy stronger parallel trend assumption: a group receiving lower `dose' of treatment should be a valid counterfactual for the group with higher `dose' (\cite{CallawayGoodmanBaconSantAnna2024_NBER32117}). In the context of catastrophic flood, less exposed districts could be historically naive to such extreme events. As they now face spillovers from statewide flood, they are less likely to be an ideal counterfactual. Moreover, continuous treatment measure or `dose' response is sensitive to functional form of continuous treatment (\cite{HiranoImbens2004_ContPS}).

I provide the descriptive statistics in Appendix \ref{app:desc-stat}. Except for women's years of education, it shows that the 
standardized mean differences between treatment and control groups are small. This implies that Tamil Nadu is valid counterfactual.

\section{Identification Strategy}\label{ident}
The goal of this paper is to examine the timing and extent of both impact and recovery from flood shock by gender. More specifically, I investigate whether men or women experience earlier disruptions in their employment and whether the recovery patterns differ by gender.
Hence, I use the event study design developed by Schythe and Clarke (2021) as specified in the following:

\begin{equation}\label{eq1}
Y_{ihdt}^{g} = \alpha^{g} +
\sum_{\substack{k=-4 \\ k \neq -1}}^{3} \delta_k^{g} (\text{Treat}_d \times \mathbf{1}\{t = k\}) +\delta_5^{g}\text{Treat}_d+
\delta_6^{g}\mathbf{X}_{idt} + \delta_7^g\mathbf{Z}_{hdt}  + \mu_d + \gamma t + \varepsilon_{ihdt}
\end{equation}

In the above empirical specification, $y_{ijdt}$ measures paid labor outcome for individual $i$ from household $h$ in district $d$ in the quarter $t$. Paid labor outcome is captured by:\\
(a) \textbf{Paid Employment:} A binary variable which takes value 1 if an individual was employed in paid jobs during the reference week, and 0 otherwise.\\
(b) \textbf{Working Hours:} A continuous variable which captures how many hours an individual spent on paid work during the last seven days. PLFS provides data on hours worked for each day of the reference week (Day 1 to Day 7). I aggregated these daily working hours to construct weekly working hours for maintaining consistency with weekly occupational status. 

$\text{Treat}_d$ is a treatment dummy which takes value 1 if individual $i$ from household $h$ is located in highly (less) exposed district $d$ and 0 if located in unaffected district from neighboring state, Tamil Nadu. $\mathbf{1}\{\text{Event}=k\}$ is a dummy variable for $k$ quarters relative to the onset of the flood. The base period is April 2018-June 2018 (i.e., $k = -1$), which is one quarter preceding the shock. 
Because the PLFS data is available during each quarter and the flood occurred in August 2018, I align the quarters with flood event as following: 
\begin{figure}[ht]
    \centering
    \includegraphics[width=0.5\linewidth]{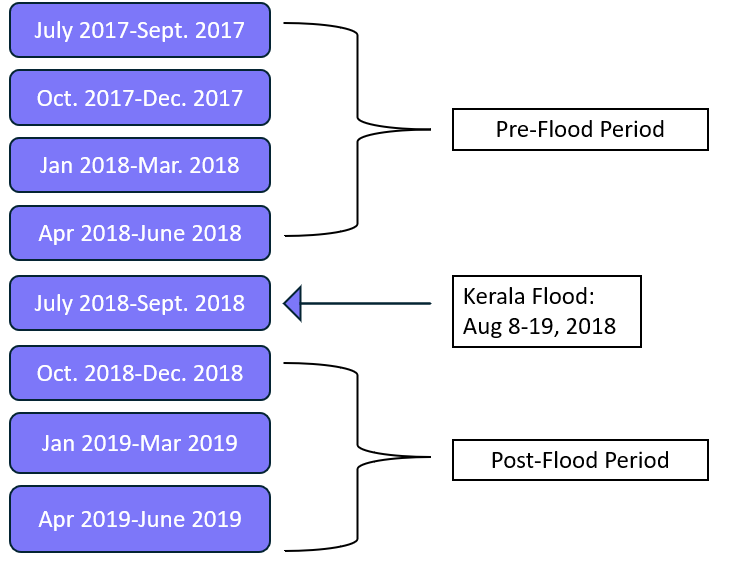}
    \caption{Timeline of Flood and Data Availability}
    \label{fig:timeline}
\end{figure}

$\mathbf{X}_{idt}$ is vector of individual level controls (age, years of education and marital status). $\mathbf{Z}_{hdt}$ is vector of household level controls (religion, social group). $\mu_d$ is district level fixed effects and $\gamma t$ captures linear time trend. $\varepsilon_{ihdt}$ is error term.

\section{Results}\label{results}
\subsection{Main Results}
To validate the results as causal estimate, it is important to test parallel trend assumption. If there exists no statistically significant differences in pre-flood labor market trends between treatment and control groups, it supports the validity of attributing the post-event differences in outcomes to the 2018 August
flood and the regression estimates can be inferred as causal. \\
I present the effect of the 2018 August flood on
paid labor outcomes separately for highly exposed districts and less exposed districts as treatment groups with unaffected districts from Tamil Nadu as control group during each quarter of July 2017-June 2019. Note that the base period is April 2018-June 2018 (${t=-1}$).

\begin{figure}[H]
    \centering
    % Left image
    \begin{subfigure}{0.48\linewidth}
        \centering
        \includegraphics[width=\linewidth]{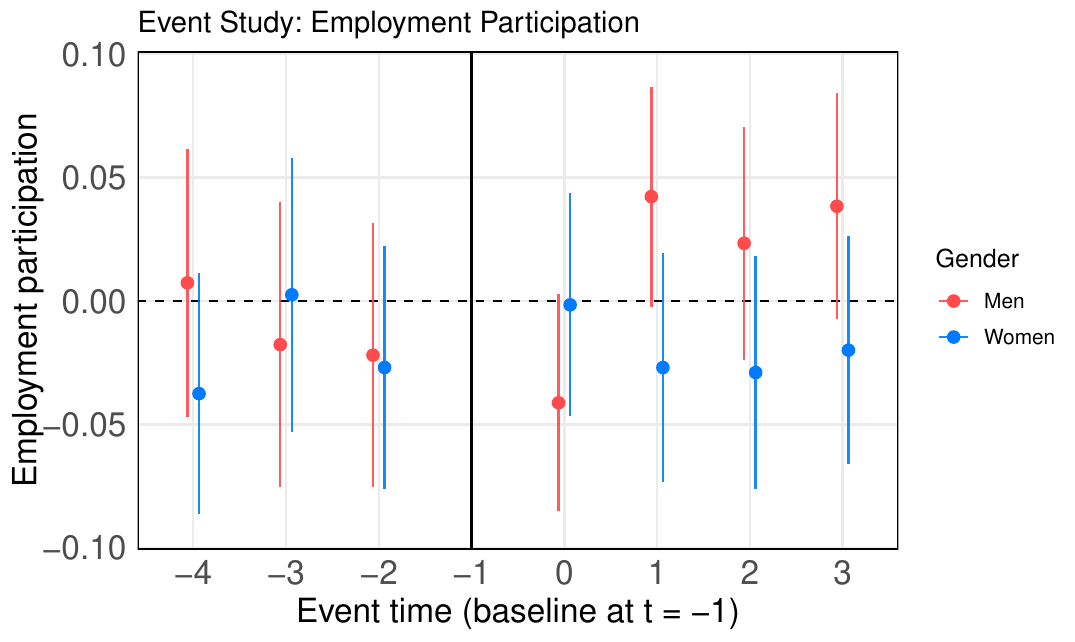}
        \caption{Event Study: Employment Participation}
        \label{fig:high1}
    \end{subfigure}
    \hfill
    % Right image
    \begin{subfigure}{0.48\linewidth}
        \centering
        \includegraphics[width=\linewidth]{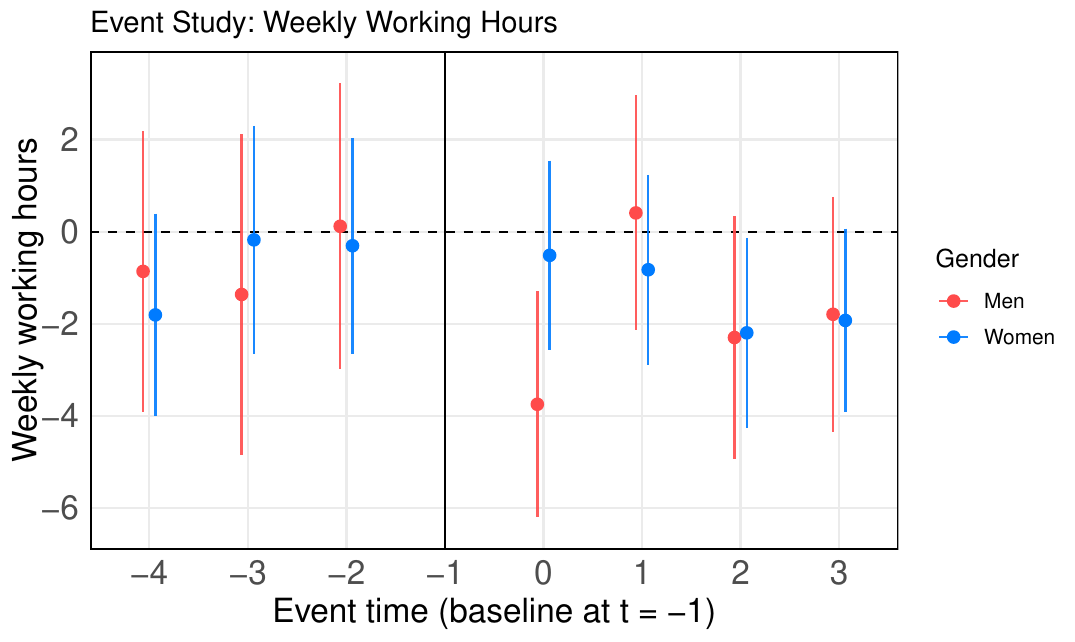}
        \caption{Event Study: Weekly Working Hours}
        \label{fig:high2}
    \end{subfigure}
    \caption{\textit{Event Study Results for Paid Labor Outcomes in highly exposed districts as treatment group. Here, the quarter April 2018- June 2018 $({t=-1})$ is used as reference period. Note: Circular dots and vertical lines show estimates and 95\% confidence intervals.}}
    \label{fig:kerala_flood1}
\end{figure}

\textbf{Highly Exposed Districts as Treatment Group:}
Figures~\ref{fig:high1} and ~\ref{fig:high2} plot the event study estimates of $\delta_k^g$ for employment participation and weekly working hours as outcome variables. In Appendix, table~\ref{tab:eventstudy_combined} gives the detail about the corresponding coefficients. Because the pre-flood estimates are statistically insignificant for both men and women across extensive and intensive margins (Figures ~\ref{fig:high1}-~\ref{fig:high2}), the parallel trend assumption holds. 

In case of extensive margin, men experienced decline in their employment participation at the time of flood in highly exposed districts (Figure~\ref{fig:high1}; Table~\ref{tab:eventstudy_combined}). They were 4.1 percentage points (pp) less likely to be employed at the time of flood compared to control group but rebounded thereafter by $\mathrm{4.2}$ pp at $t=1$ a (Figure~\ref{fig:high1}; Column (2), Table~\ref{tab:eventstudy_combined}). Women also witnessed reduction in their employment participation (by $\mathrm{1.9- 2.7}$pp) in highly exposed districts but not statistically insignificant (Figure~\ref{fig:high1}; Columns (3)- (4), Table~\ref{tab:eventstudy_combined}). This can be partly attributable to their low participation during the baseline period. With less employed women (i.e., many 0’s and few 1’s), there is less scope to significantly observe additional exits from labor market. Hence, the binary paid labor measurement is less sensitive to small changes.
In case of intensive margin, men’s paid labor fell by $\mathrm{3.7}$ hours per week when flood struck in highly exposed districts but recovered subsequently (Figure~\ref{fig:high2}; Column (2), Table~\ref{tab:eventstudy_combined}). Unlike men, women witnessed gradual but persistent disruptions: women faced reduction in their paid labor by $\mathrm{2.2}$ hours per week and $\mathrm{1.9}$ hours per week in highly exposed districts at $t=2$ and $t=3$ respectively (Figure~\ref{fig:high2}; Column (4), Table~\ref{tab:eventstudy_combined}). Women's delayed recovery could be attributable to two factors: first, working women are more concentrated in-service sectors which depend on restoration of infrastructure facilities. And the reopening of workplaces takes time after catastrophic weather event (later shown in \ref{hetero}); second, women spend increased time on domestic and caregiving responsibilities after flood (Sarkar 2024) and hence, those who are employed will reduce their paid labor hours.

\begin{figure}[ht]
    \centering
    % Left image
    \begin{subfigure}{0.48\linewidth}
        \centering
        \includegraphics[width=\linewidth]{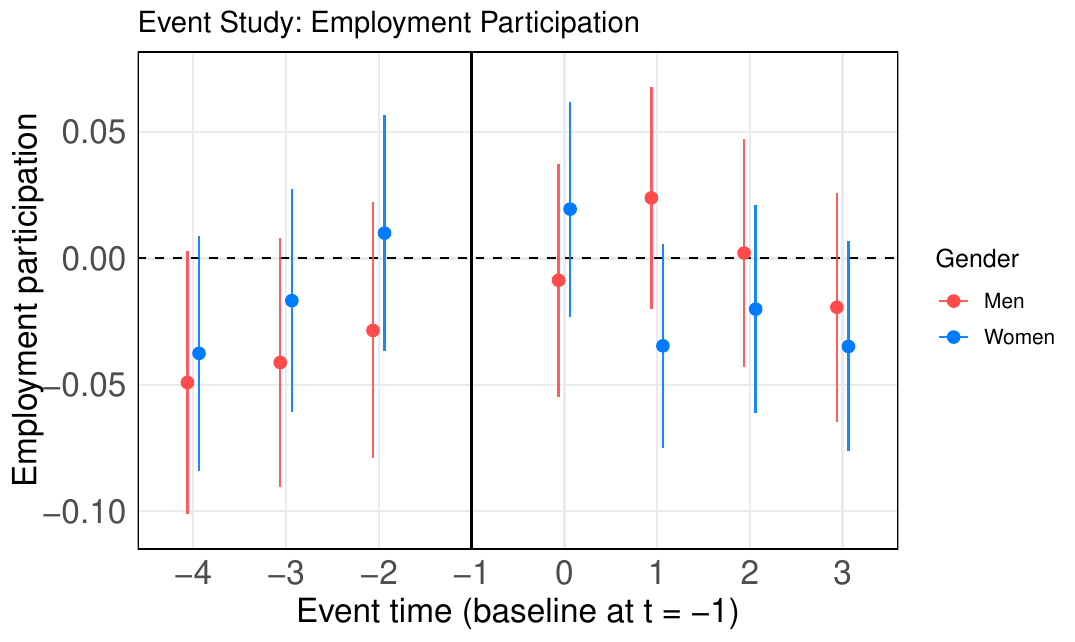}
        \caption{Event Study: Employment Participation}
        \label{fig:low1}
    \end{subfigure}
    \hfill
    % Right image
    \begin{subfigure}{0.48\linewidth}
        \centering
        \includegraphics[width=\linewidth]{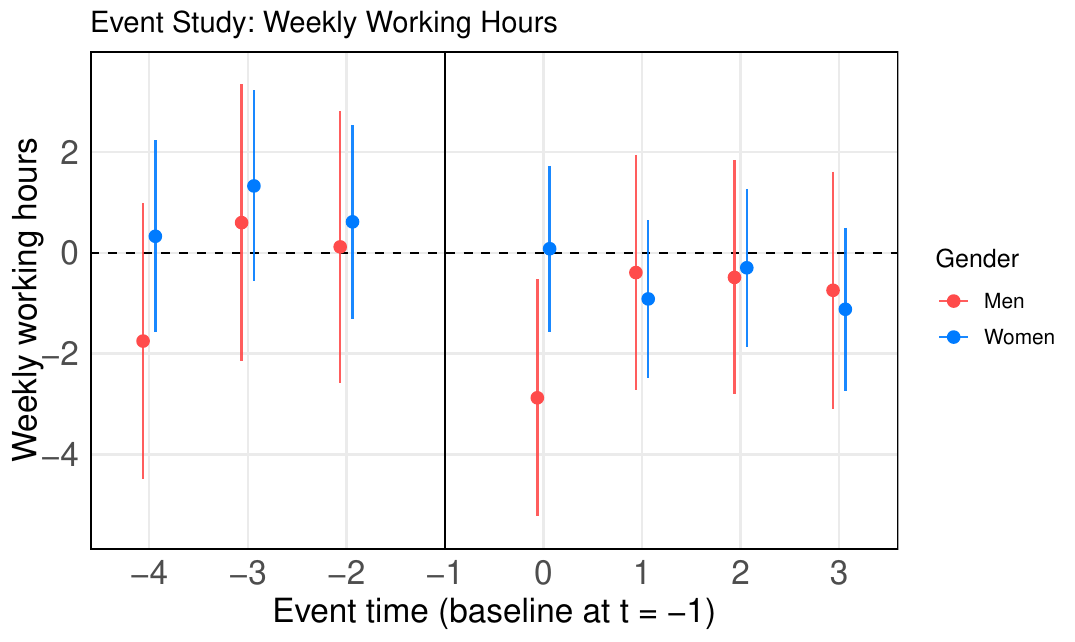}
        \caption{Event Study: Weekly Working Hours}
        \label{fig:low3}
    \end{subfigure}
     \caption{\textit{Event Study Results for Paid Labor Outcomes in less exposed districts as treatment group. Here, the quarter April 2018-June 2018 $(t=-1)$ is used as reference period. Note: Circular dots and vertical lines show estimates and 95\% confidence intervals.}}
    \label{fig:kerala_flood2}
\end{figure}

\textbf{Less Exposed Districts as Treatment Group:} Figures ~\ref{fig:low1} and ~\ref{fig:low3} plot the event study estimates of $\delta_k^g$ for employment participation and weekly working hours as outcome variables; Table~\ref{tab:lessaffected_combined} gives the detail about the corresponding coefficients. 
For women, the parallel trend assumption is satisfied as the pre-flood estimates are statistically insignificant (Figures~\ref{fig:low1}- \ref{fig:low3}; Column (4), Table~\ref{tab:lessaffected_combined}). While pre-flood trends in working hours are also statistically insignificant for men, there exists pre-flood differences in men's employment participation at ${t=-4}$ (Figure~\ref{fig:low1}; Columns (1)- (2), Table~\ref{tab:lessaffected_combined}).

In case of extensive margin, men's employment participation was statistically insignificant in less affected districts after flood. On the contrary, women experienced brief disruption by $\mathrm{3.5}$ pp at ${t=1}$ in less affected districts compared to the comparison group but they recovered thereafter (Figure~\ref{fig:low1}; Column (4), Table~\ref{tab:lessaffected_combined}). On intensive margin, men faced decline in their paid labor by $\mathrm{2.87}$ hours per week in less affected districts at the time of flood (Figure~\ref{fig:low3};  Column (2), Table~\ref{tab:lessaffected_combined}) but rebounded from ${t=1}$ onwards. In less affected districts, women did not see any statistically significant drop in their weekly working hours compared to their control group.

These results indicate that men in highly exposed districts faced short-lived labor market disruption while women experienced gradual and persistent decline in their paid labor hours. On the contrary, both men and women in less exposed districts witnessed only temporary reduction in their paid labor outcomes. Hence, I focus my remaining analysis on highly affected areas as treatment group.
\subsection{Robustness Checks}

To test the robustness of the main findings, I estimate \eqref{eqa} using alternative treatment definitions while the control group remain unchanged. These alternative specifications help assess whether the observed labor market effects of 2018 flood were concentrated in the most affected areas.

\subsection{Alternative Threshold}\label{sec:robust:p75}
\begin{figure}[h]
    \centering
    \begin{subfigure}{0.48\linewidth}
        \centering
        \includegraphics[width=\linewidth]{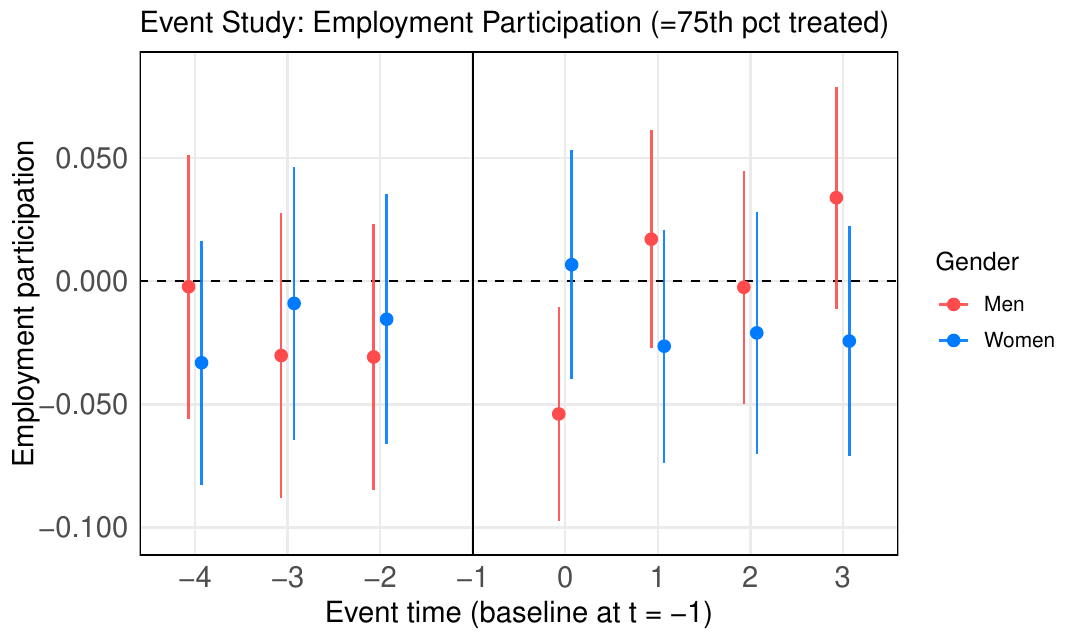}
        \caption{Event Study: Employment Participation}
        \label{fig:rob_7}
    \end{subfigure}
    \hfill
    \begin{subfigure}{0.48\linewidth}
        \centering
        \includegraphics[width=\linewidth]{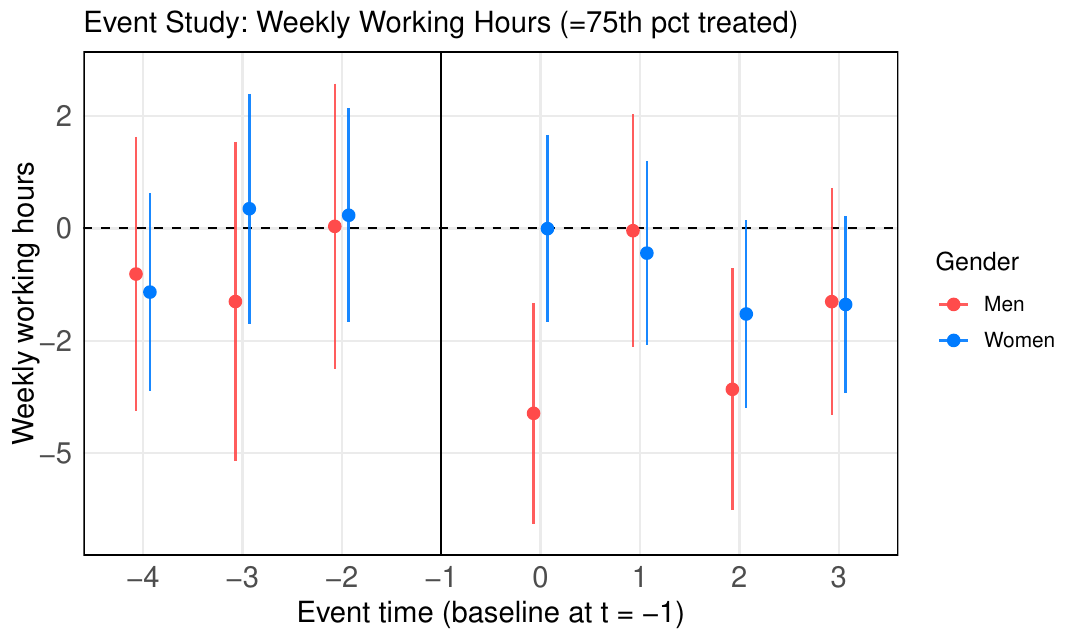}
        \caption{Event Study: Weekly Working Hours}
        \label{fig:rob_8}
    \end{subfigure}
    \caption{\textit{Event Study Results for Paid Labor Outcomes in districts with high flood-induced landslides as treatment group. Here, the quarter April 2018- June 2018 $({t=-1})$ is used as reference period. Note: Circular dots and vertical lines show estimates and 95\% confidence intervals.}}
\end{figure}

I redefine treatment group as those Kerala districts with flood exposure at 75$^\text{th}$ percentile and above. Event–time estimates for employment participation and weekly working hours are shown in Figure~\ref{fig:rob_7} and \ref{fig:rob_8}. Details about estimated coefficients are in table~\ref{tab:robust_p75_full}. 

On extensive margin, men were 5.4 pp less likely to be employed in treated districts at $t=0$ but recovered thereafter (Figure~\ref{fig:rob_7}; Column (2), Panel A, Table~\ref{tab:robust_p75_full}). But women's employment participation also declined during post flood period but was statistically insignificant (Figure~\ref{fig:rob_7}; Column (4), Panel A, Table~\ref{tab:robust_p75_full}). On intensive margin, men experienced decline in their paid labor by 4.11 hours per week at $t=0$ and $t=2$ but rebounded at $t=3$ (Figure~\ref{fig:rob_8}; Column (2), Panel B, Table~\ref{tab:robust_p75_full}). Women, on the contrary, experienced reduction with a lag: they faced decline in paid labor by $\mathrm{1.90}$ hours per week and $\mathrm{1.69}$ hours per week at $t=2$ and $t=3$ respectively(Figure~\ref{fig:rob_8}; Column (4), Panel B, Table~\ref{tab:robust_p75_full}).
Thus, the results are consistent with key findings even when threshold is tightened.

\subsection{Spatial Spillovers}
\begin{figure}[H]
    \centering
    \begin{subfigure}{0.48\linewidth}
        \centering
        \includegraphics[width=\linewidth]{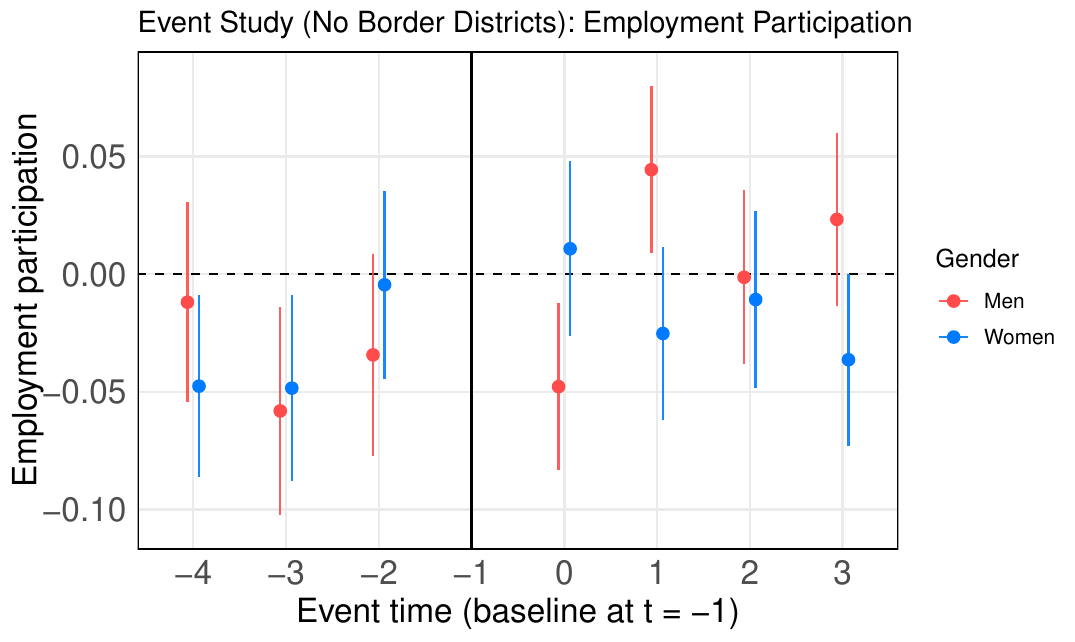}
        \caption{Event Study: Employment Participation}
        \label{fig:rob_3a}
    \end{subfigure}
    \hfill
    \begin{subfigure}{0.48\linewidth}
        \centering
        \includegraphics[width=\linewidth]{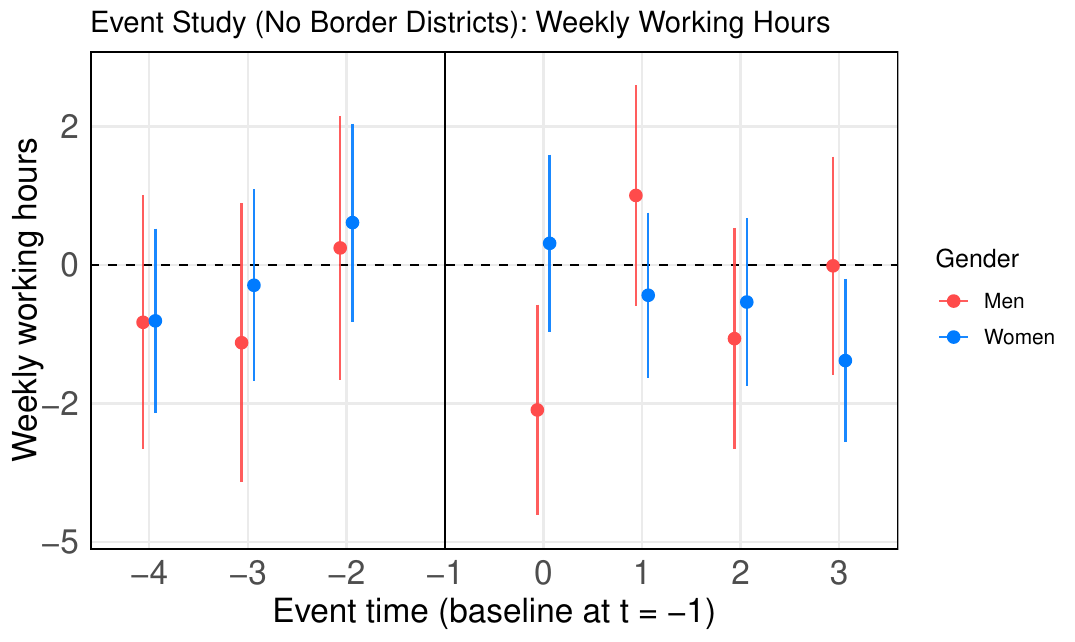}
        \caption{Event Study: Weekly Working Hours}
        \label{fig:rob_3b}
    \end{subfigure}
    \caption{\textit{Event Study Results for Paid Labor Outcomes by dropping border districts. Here, the quarter April 2018- June 2018 $({t=-1})$ is used as reference period. Note: Circular dots and vertical lines show estimates and 95\% confidence intervals.}}
\end{figure}

It is possible that people living close to the state border could temporarily migrate during flood shock. Because the PLFS data gives geographical location of individuals and households at district level only, so I exclude Kerala districts sharing border with unaffected states from the analysis. Results are shown in figure~\ref{fig:rob_3a}--\ref{fig:rob_3b}. Details about estimates are in table~\ref{tab:spillover_robust}. 

On extensive margin, men faced fell in their employment participation by $\mathrm{4.8}$ pp during $t=0$ but rose by $\mathrm{4.4}$ pp during $t=1$ (Figure~\ref{fig:rob_3a}; Panel A,  Column (2), Table~\ref{tab:spillover_robust}). On the contrary, women's employment participation did not change significantly after flood (Figure~\ref{fig:rob_3b}; Panel A,  Column (2), Table~\ref{tab:spillover_robust}). On intensive margin, men experienced decline in their employment by $\mathrm{2.6}$ hours per week during $t=0$ but rebounded thereafter (Figure~\ref{fig:rob_3b}; Panel B,  Column (2), Table~\ref{tab:spillover_robust}). Women saw reduction in their working hours with lag: their paid labor declined by 1.7 hours per week (Figure~\ref{fig:rob_3b}; Panel B,  Column (4), Table~\ref{tab:spillover_robust}). These results follow similar trends as the main findings during post-flood periods. 

However, it should be noted that there exists significant pre-trends at $t=-3$ for men and $t=-4$ for women in case employment participation outcome. Hence, the corresponding coefficients should be interpreted with caution. Nevertheless, the post-flood trends in employment participation are similar to key findings.

\subsection{Heterogeneous Effects}\label{hetero}
I estimate \eqref{eq1} separately for supply and demand side factors. On the supply side, I test how results differ by social norms, specifically, by marital status and dependency burden. On the demand side, I examine how results vary by  employment in different types of industries. Complete tables with estimates are in the Appendix.
\subsubsection{Social Norms}
Do social norms emanating from gender difference influence men’s and women’s paid labor responses? Gendered norms around home production and caregiving responsibilities are likely to reduce women’s paid labor. This is most likely to be the case for married women and those who have dependent family members (i.e., children and elderly). On the contrary, society expects men’s responsibilities in terms of earning by working outside. This is likely to be more pronounced when they have more mouths to feed. 

\begin{figure}[ht]
    \centering
    \includegraphics[width=0.9\linewidth]{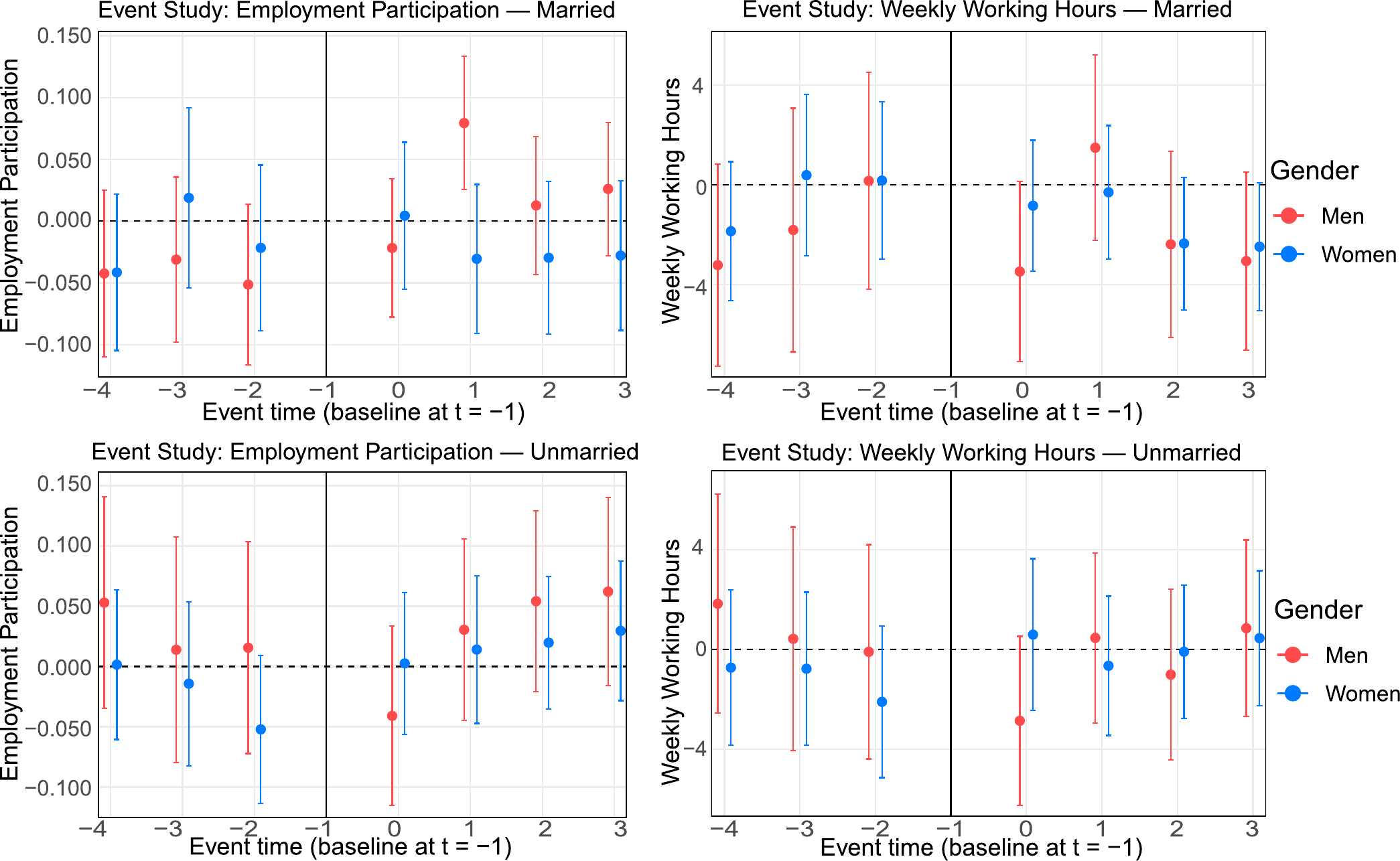}
    \caption{\textit{Heterogeneous Effects of Flood by Marital Status. Upper panel gives results for married men and women. Lower panel reports results for unmarried men and women. Here, the quarter April 2018- June 2018 $({t=-1})$ is used as reference period. Note: Circular dots and vertical lines show estimates and 95\% confidence intervals.}}
    \label{fig:married}
\end{figure}

\textbf{(a) Marital Status.} In figure~\ref{fig:married}, the upper two panels show results for both extensive and intensive margin of married men and women while the lower two panels give the same estimates for unmarried men and women. Details about estimated coefficients are in Panels (A) and (B) of table~\ref{tab:eventstudy_marital}.

Married men were $\mathrm{7.9}$ pp more likely to be employed at $t=1$ but married women were $\mathrm{3.0}$ pp less likely to be employed at ${t=1}$ and ${t=2}$, though not statistically significant (Figure~\ref{fig:married}; Panel (A), Columns (2) and (4), Table~\ref{tab:eventstudy_marital}). Unmarried men and women exhibit statistically insignificant increase in their employment participation after flood (Figure~\ref{fig:married}; Panel (A), Columns (1) and (3), Table~\ref{tab:eventstudy_marital}).

On the intensive margin (weekly paid hours), unmarried men faced decline in their working hours by $\mathrm{2.86}$ hours per week when the flood occurred $({t=0})$ but unmarried women did not see any reduction in their working hours after flood (Figure~\ref{fig:married}; Panel (B), Columns (1) and (3), Table~\ref{tab:eventstudy_marital}). On the contrary, both married men and women experienced decline in their paid labor hours. Married men spent $\mathrm{3.5}$ hours per week and $\mathrm{3}$ hours per week less in their paid jobs at ${t=0}$ and ${t=3}$ (Figure~\ref{fig:married}; Panel (B), Column (2), Table~\ref{tab:eventstudy_marital}). Married women spent $\mathrm{2.36}$ hours per week and 2.5 hours per week less in employment at ${t=2}$ and ${t=3}$ (Figure~\ref{fig:married}; Panel (B), Column(4), Table~\ref{tab:eventstudy_marital}).

Thus, the flood shock reallocates paid labor based on their marital status. Unmarried women did not see any significant change in their employment. Married women faced substantial decline their paid labor. This could be attributable to their increased care burden. As flood hit, children are more likely to stay home due to school closures and both elderly and children are more vulnerable to water-borne diseases after flood (\cite{AggarwalEtAl2025}; \cite{DimitrovaMuttarak2020}). Married men buffer financial loss by quick re-entry but spend lesser hours. The latter is attributable to flood-induced disruptions in infrastructure and physical capital.

\textbf{(b) Dependency Ratio.} One limitation of PLFS data is that it does not link household members in a way that would help me to identify one-to-one relationship between household members (e.g., whether an individual has children, how many children or whether they live with their parents or parent-in-law). But the presence of dependent population shapes gender differences in paid labor supply (e.g., Maestas et al.\ 2023, Afridi et al.\ 2022). To capture this, I define dependency ratio at household level as the share of household members who are children (0- 14 years) and elderly (65 years \& above):
\begin{equation}
\text{Dependency Ratio}
= \frac{\#\{\text{Household members aged }0\text{–}14\}
\;+\; \#\{\text{Household members aged }65+\}}
{\text{Household Size}} \, .
\end{equation}

In India, individuals aged above 14 years are eligible to work in non-hazardous jobs and most people usually retire around 60- 65 years. I classify households as high dependency if the dependency ratio is above median and low dependency otherwise. In figure 8, the upper two panels show results for both extensive and intensive margin of men and women among households with high dependency while the lower two panels give the same estimates for those in households with low dependency.

\begin{figure}
    \centering
    \includegraphics[width=1.0\linewidth]{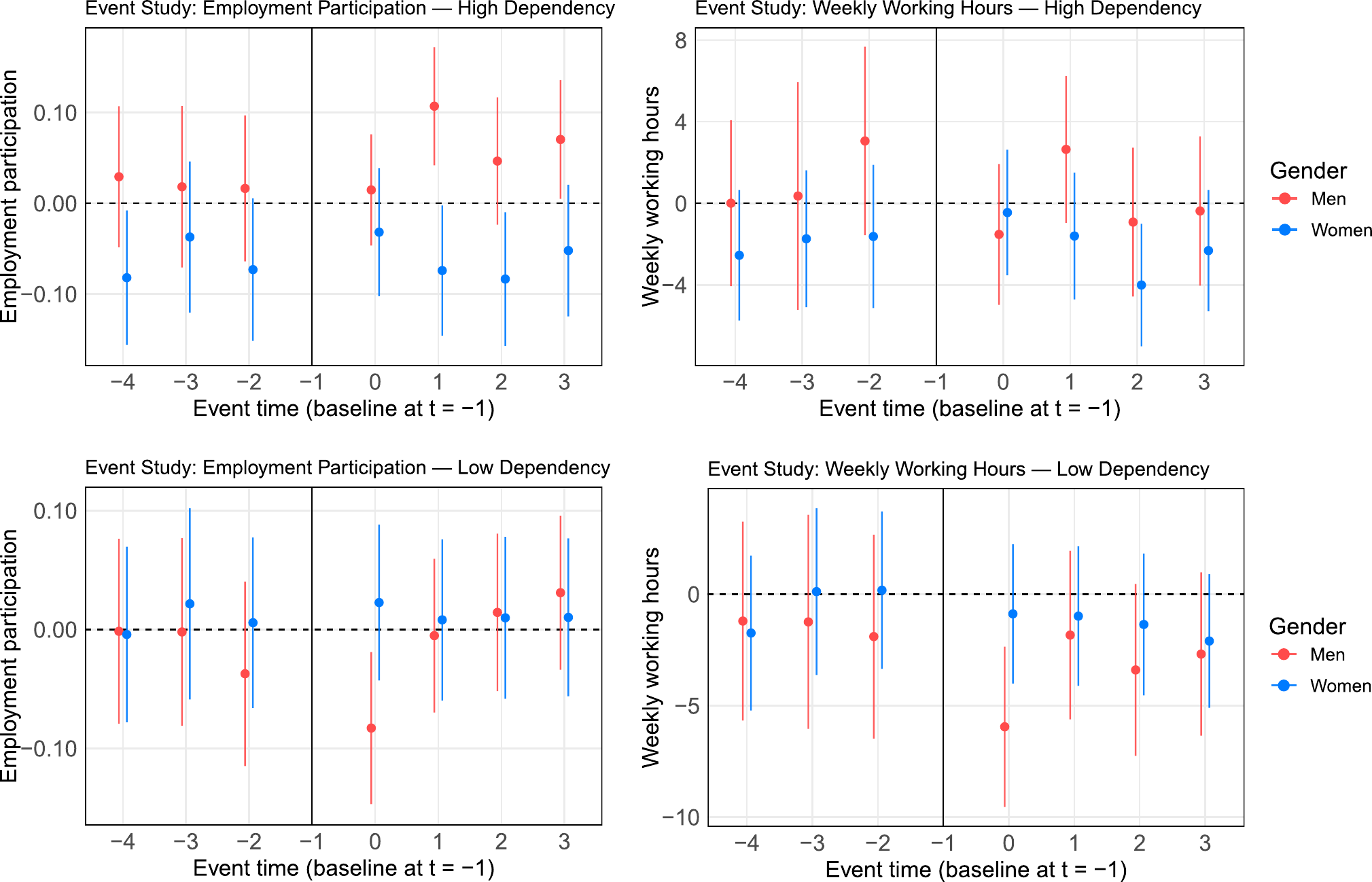}
    \caption{\textit{Heterogeneous Effects of Flood by Dependency Burden. Upper panel presents results for men and women coming from households with high dependency ratio. Lower panel reports results for men and women from households with low dependency ratio. Here, the quarter April 2018- June 2018 $({t=-1})$ is used as reference period. Note: Circular dots and vertical lines show estimates and 95\% confidence intervals.}}
    \label{fig:dep}
\end{figure}

In high dependency households, women saw decline in their employment participation by $\mathrm{7.4}$ pp and $\mathrm{8.3}$ pp at ${t=1}$ and ${t=2}$ while men were $\mathrm{10.7}$ pp and $\mathrm{7}$ pp more likely to be employed at ${t=1}$ and ${t=3}$ (Figure~\ref{fig:dep}; Panel A, Columns (2) and (4), Table~\ref{tab:eventstudy_depburden}). In low dependency households, on the contrary, the flood had no significant effect on women's employment participation while men experienced decline in their employment participation by $\mathrm{8.3}$ pp when the flood hit ($({t=0)}$ (Figure~\ref{fig:dep}; Panel A, Columns (1) and (3), Table~\ref{tab:eventstudy_depburden}). 

Panel B also shows results for intensive margin. In high dependency households, women reduced their paid labor by 4 hours per week at $(\mathrm{t=2})$ (Figure~\ref{fig:dep}; Column (4), Panel B, Table~\ref{tab:eventstudy_depburden}) while men had increased their paid labor by 2.65 hours per week at $(\mathrm{t=1})$, though not statistically significant (Figure~\ref{fig:dep}; Column (2), Panel B, Table~\ref{tab:eventstudy_depburden}). In low-dependency households, women had no significant change in their paid labor hours per week while men faced reduction in their paid labor hours by 5.95 hours per week at $(\mathrm{t=0})$ (Figure~\ref{fig:dep}; Columns (1) and (3), Panel B, Table~\ref{tab:eventstudy_depburden}). 

Thus, women tend to cut down on their paid labor when faced with higher dependency while men increase their participation in paid jobs under similar circumstance. 

\subsubsection{Employment by Industry}

\begin{figure}[ht]
    \centering
    \includegraphics[width=0.8\linewidth]{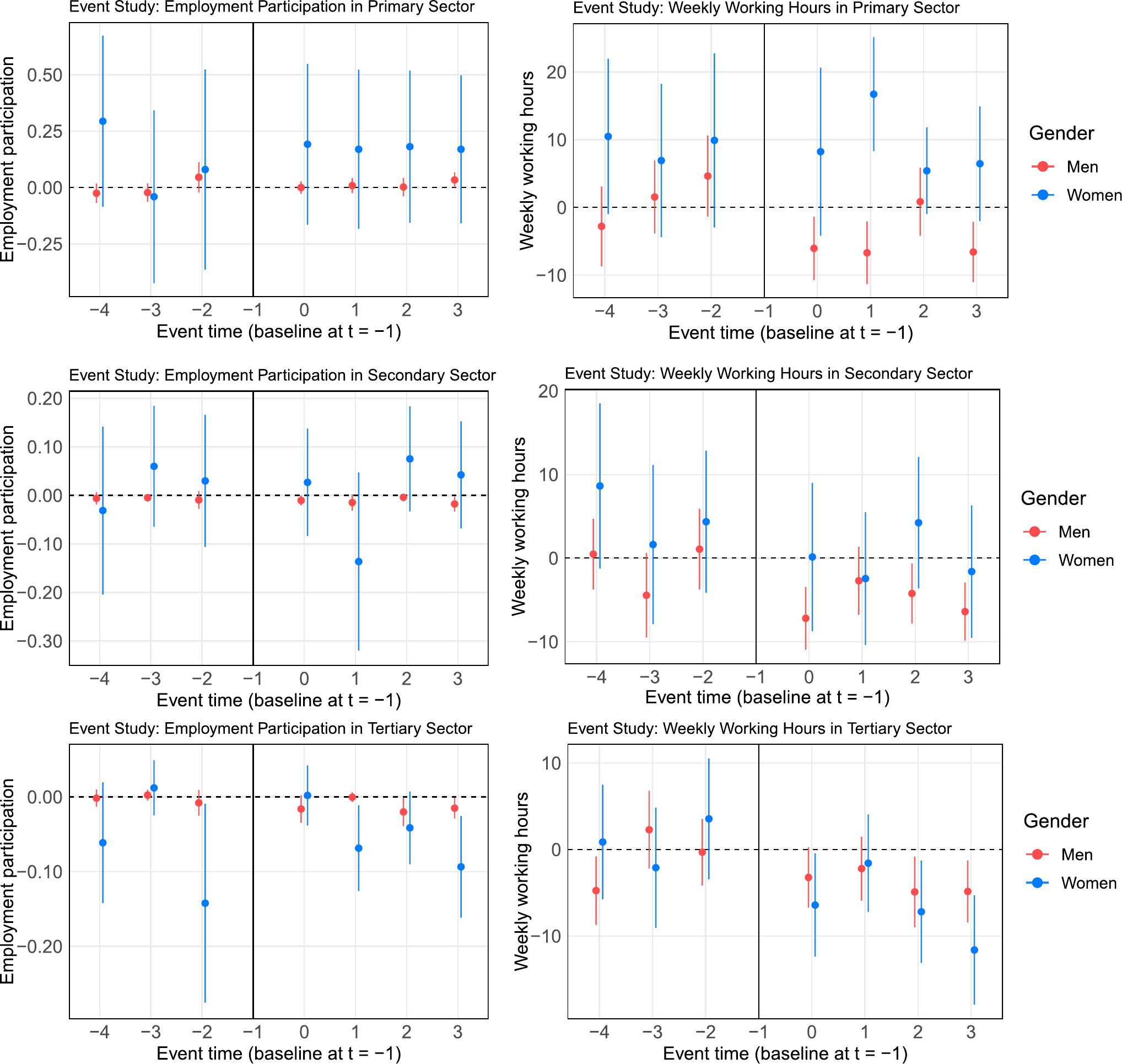}
    \caption{\textit{Heterogeneous Effects of Flood by Industry. Here, the quarter April 2018- June 2018 $({t=-1})$ is used as reference period. Top, middle and lower panels respectively present results for primary, secondary and tertiary sector. Note: Circular dots and vertical lines show estimates and 95\% confidence intervals.}}
    \label{fig:fig_industry}
\end{figure}

\textbf{(a) Primary sector.}
The primary sector consists of (i) agriculture and allied activities and (ii) mining and quarrying. In Figure~\ref{fig:fig_industry}, the top panel shows the event study results for paid labor of men and women in primary sector. Columns (1) and (4) in table~\ref{tab:eventstudy_bysector} presents the corresponding coefficients and standard errors. \\
On the extensive margin, women's employment participation did not significantly change after flood (Figure~\ref{fig:fig_industry}; Panel A, Column (4), Table~\ref{tab:eventstudy_bysector}). Men saw negligible change in their participation up to $t=2$ but increased by nearly 3.4 pp at $t=3$ (Figure~\ref{fig:fig_industry}; Panel A, Column (1), Table~\ref{tab:eventstudy_bysector}). However, this did not translate into higher working days per week. On the intensive margin, men faced consistent decline by $\mathrm{6.0- 6.7}$ hours per week during $t=0-  t=3$ (Figure~\ref{fig:fig_industry}; Panel B, Column (1), Table~\ref{tab:eventstudy_bysector}). But women spent $\mathrm{16.71}$ hours per week and $\mathrm{5.4}$ hours per week more in primary sector at $t=1$ and $t=2$ respectively (Figure~\ref{fig:fig_industry}; Panel B, Column (4), Table~\ref{tab:eventstudy_bysector}). While pre-trend does exist for women's working hours at $t=-4$, causal interpretation should be done cautiously. Despite this, the flood is associated with women's shift in primary sector while men re-later.

\textbf{(b) Secondary sector.}
This sector includes (i) manufacturing, (ii) construction, (iii) water supply, sewerage, waste management and remediation activities, and (iv) electricity, gas, steam and air conditioning supply. In Figure~\ref{fig:fig_industry}, the middle panel shows the event study results for paid labor of men and women in secondary sector. Columns (2) and (5) in table~\ref{tab:eventstudy_bysector} presents the corresponding coefficients and standard errors. \\ 
In case of extensive margin, men were $\mathrm{1.07}$ pp less likely to be employed when the flood hit ${(t=0)}$ and continue to fall by $\mathrm{1.49}$- $\mathrm{1.79}$ pp during ${t=1} - {t=3}$ (Figure~\ref{fig:fig_industry}; Panel A, Column (2), Table~\ref{tab:eventstudy_bysector}). On the contrary, women's employment participation did not face any significant change in this sector after flood (Figure~\ref{fig:fig_industry}; Panel A, Column (5), Table~\ref{tab:eventstudy_bysector}). Similar trends are observed in case of intensive margin: men spent $\mathrm{4.3- 7.2}$ hours per week less in secondary sector at ${t=0}$, ${t=2}$ and ${t=3}$ while women's paid labor hours did not show any discernible change after flood (Figure~\ref{fig:fig_industry}; Panel A, Columns (2) and (5), Table~\ref{tab:eventstudy_bysector}). However, there exists pre-trend for women's paid labor hours at $t=-4$. Overall, men experienced substantial setback in secondary sector and recover slowly but women showed very little change. 
 
\textbf{(c) Tertiary sector.}
This sector covers (i) wholesale and retail trade, (ii) transportation and storage, (iii) information and communication, and other service activities\footnote{e.g., accommodation and food service; financial and insurance; real estate; professional, scientific and technical activities.}. In Figure~\ref{fig:fig_industry}, the lower panel shows the event study results for paid labor of men and women in secondary sector. Columns (3) and (6) in table~\ref{tab:eventstudy_bysector} presents the corresponding coefficients and standard errors. \\
On the extensive margin, men were $\mathrm{1.6}$ pp less likely to be employed when the flood occurred ${t=0}$ and continue to remain lower by $\mathrm{1.5- 2.0}$ pp during ${t=2}- {t=3}$ (Figure~\ref{fig:fig_industry}; Panel A, Column (3), Table~\ref{tab:eventstudy_bysector}). Women experienced delay yet substantial loss in employment participation: they were $\mathrm{6.9}$ pp less likely to be employed at $t=1$ and gradually fell further by $\mathrm{4.15}$ pp at ${t=2}$ and by $\mathrm{9.4}$ pp at ${t=3}$ (Figure~\ref{fig:fig_industry}; Panel A, Column (6), Table~\ref{tab:eventstudy_bysector}). I find similar effects in case of intensive margin of labor market outcomes. Men lost $\mathrm{3.2- 4.9}$ paid labor hours per week during $\mathrm{t=0}- \mathrm{t=3}$ (Figure~\ref{fig:fig_industry}; Panel B, Column (3), Table~\ref{tab:eventstudy_bysector}). Women experienced much larger loss: their paid labor declined by $\mathrm{6.4- 11.6}$ hours per week during $\mathrm{t=0}- \mathrm{t=3}$ (Figure~\ref{fig:fig_industry}; Panel B, Column (6), Table~\ref{tab:eventstudy_bysector}). These estimates show that tertiary sector was worst affected by flood. Note that a significant pre-trend exists at $t=-4$ for men's employment in tertiary sector  (Figure~\ref{fig:fig_industry}; Panel A, Column (6), Table~\ref{tab:eventstudy_bysector}). Nevertheless, both men and women faced decline in their paid labor hours but the contractions are much larger for women in this sector.

Thus, women were hit hardest in tertiary sector where their working hours did not recover even 9 months after the flood occurred. In case of men, they faced highest loss in secondary sector as their paid labor hours were slow to recover in this sector. While men did increase their participation in primary sector by the end of 6 months since flood, this did not translate into more labor hours. Secondary and tertiary sector fared worst due to widespread disruptions in infrastructure facilities (including airports) and transport services. And primary sector depends mostly on natural resources and require less commuting unlike secondary and tertiary sectors.

\newpage
\section{Conclusion}\label{conclude}

I estimate the gender-specific dynamic effects of a catastrophic coastal flooding on labor market. My key results show that men faced immediate decline in their paid labor when flood hit, while women experienced delayed reduction. The dynamic estimates also suggest that men rebounded quickly while women's recovery was slow. These results can help in improving both the timing and targeting of disaster response actions. Main findings are governed by demand side and supply side factors. 

On the demand side, men fared worst in secondary sector. They also witnessed decline in employment in tertiary sector but recovered in six months after the flood. However, women faced persistent and larger loss in both employment and working hours in tertiary sector. They shifted their participation and time more toward primary sector. On the supply side, marital status and household dependency shape gender-differentiated paid labor outcomes. Married men cushioned flood-induced loss by re-entering the market soon after flood but married women faced contraction in their paid labor. The latter is possibly due to increased unpaid care responsibilities. In households with higher share of children and older members, men increased their participation in paid jobs while women were less likely to be employed and cut back on working hours. This aligns with gendered social norms which assign men to be the primary earning member whereas women are expected to be primary caregiver. 

As extreme weather events become more frequent and severe, policymakers must adopt gender-sensitive approaches in disaster response and economic recovery strategies. Faster restoration of infrastructure and stable conditions is important to combat loss in production of service sector activities. Safe transport, emergency childcare support and government employment programs can deter from reinforcing social norms and gender inequality in labor market. Targeted financial support to restore and repair physical capital in primary sector can address men's reduction in working hours.

Future work could examine how flood intensity impact job matching and how the pace of post-flood transport reopening affect gender inequality in labor market. These are important because liquidity and mobility constraints can lead to poor job match and reinforce gender disparities. 

\newpage
\printbibliography

\newpage
\appendix
\section{Map of Kerala}
\begin{figure}[ht]
    \centering
    \includegraphics[width=0.5\linewidth]{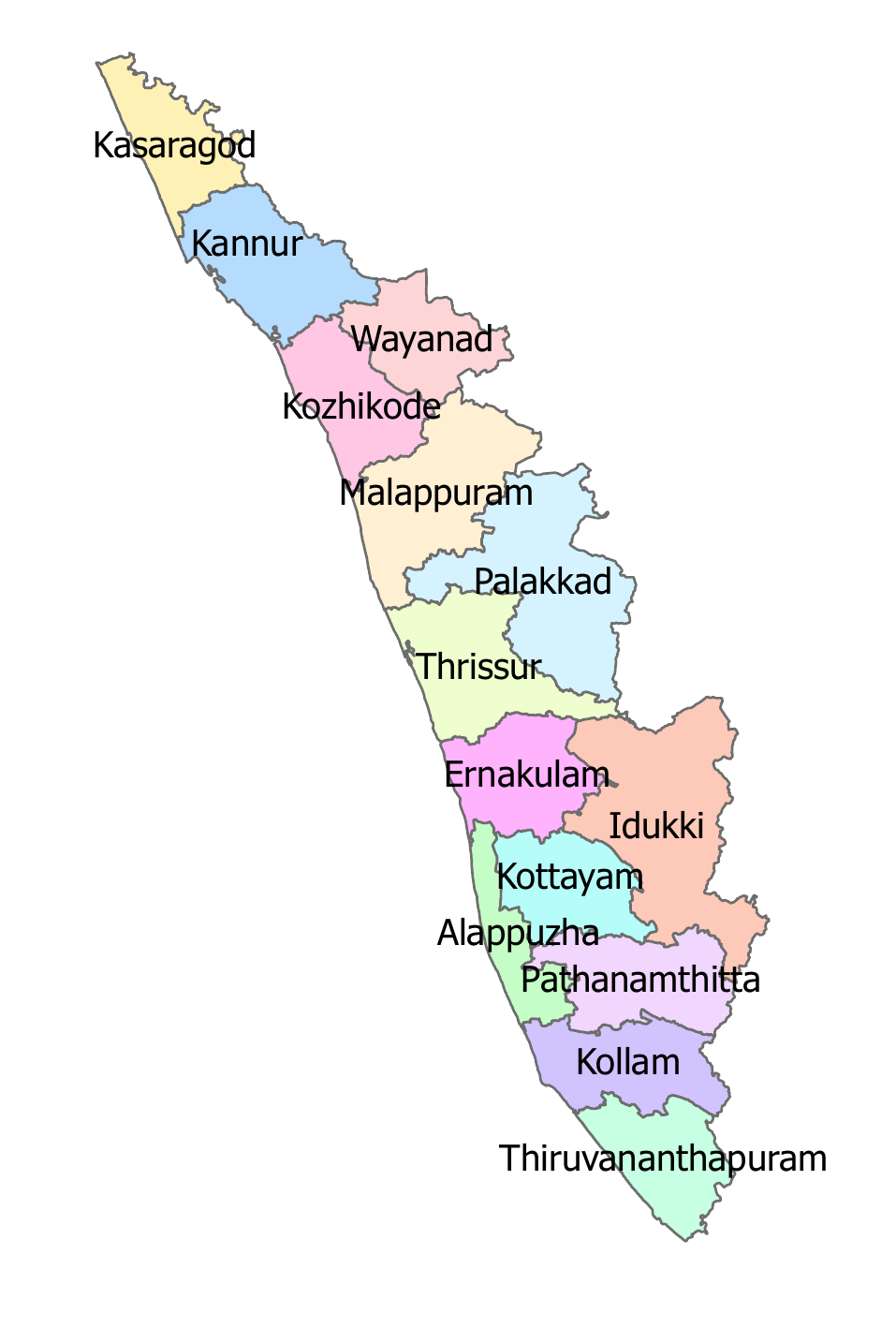}
    \caption{\textit{District Boundary of Kerala.
    Source: Self-compiled using shapefiles from \href{https://github.com/geohacker/kerala/tree/master/shapefiles}{GitHub repository}.}}
    \label{fig:kerala}
\end{figure}
\newpage
\section{Sentinel-1 SAR–derived Inundated areas: Details}
\label{subsec:inundation}

Data on 2018 flood inundation in Kerala is obtained from Sentinel-1 (VV polarization) using C-band SAR backscatter. All processing is performed on Google Earth Engine (GEE). The workflow proceeds as follows:

\textbf{Study area:}
Districts are defined from FAO GAUL level-2 for India (Kerala subset), and taluk (ADM3) boundaries are extracted from Kerala GIS asset.

\textbf{SAR image preparation:}
I assemble a pre-flood composite for March 2018 and a flood-period composite for August 2018 using \texttt{COPERNICUS/S1\_GRD}:
(i) restrict to Interferometric Wide (IW) mode; (ii) require VV polarization; (iii) include both ascending and descending passes; (iv) clip to the Kerala districts. I take the temporal \emph{median} over 2018-03-01 to 2018-03-31 for the pre-flood surface. Since the flood hit in  August, I build an August collection (2018-08-01 to 2018-09-01) for the flood surface and take the temporal \emph{minimum} to emphasize the strongest water-induced backscatter decreases during peak inundation. To suppress speckle, I apply a small spatial smoothing kernel to each composite.

\textbf{Backscatter differencing:} I compute a difference image
\begin{equation}
\Delta \text{VV} = \text{VV}_{\text{pre-flood (Mar)}} - \text{VV}_{\text{flood (Aug min)}},
\end{equation}

so that larger positive values indicate stronger backscatter drops consistent with open water or waterlogged surfaces.

\textbf{Automatic thresholding (Otsu):}
Let $h(b)$ denote the histogram of $\Delta \text{VV}$ over Kerala (10\,m scale). I use Otsu’s method to choose a threshold $t^\ast$ that maximizes between-class variance,
\begin{equation}
t^\ast = \arg\max_{t}\; \omega_0(t)\,\omega_1(t)\,\big(\mu_0(t)-\mu_1(t)\big)^2,
\end{equation}
where $\omega_k$ and $\mu_k$ are class weights and means for $\Delta \text{VV}$ below/above $t$. Pixels with $\Delta \text{VV} > t^\ast$ are provisionally labeled as flooded.

To reduce false positives from permanent water bodies, I intersect with the JRC Global Surface Water occurrence layer and mask out locations with long-run water occurrence $\geq 10\%$.\footnote{JRC Global Surface Water v1.3, GEE ID \texttt{JRC/GSW1\_3/GlobalSurfaceWater}, using band \texttt{occurrence}.} The resulting raster is our ``clean'' flood mask for August 2018. I multiply the binary flood mask by pixel area to obtain flooded hectares and summarize by polygons using zonal reduction:
(i) district-level flooded area (hectares) over GAUL L2 districts; and
(ii) taluk-level flooded area and percent inundated, computed as flooded hectares divided by taluk area. Both tables are exported from GEE as CSV files for integration with the microdata.

\section{Main District Level Flood Exposure in Kerala}
\begin{table}[!htbp]\centering
\scriptsize
\caption{Weighted flood exposure of each district in Kerala}
\label{tab:district_exposure}
\begin{threeparttable}
\setlength{\tabcolsep}{8pt}
\begin{tabular}{r l r}
\toprule
\textbf{District Code} & \textbf{District Name} & \textbf{Exposure} \\
\midrule
11 & Alappuzha       & 752.2243241 \\
 9 & Idukki          & 85.28972543 \\
 2 & Kannur          & 40.26112356 \\
 1 & Kasaragod       & 48.28403049 \\
13 & Kollam          & 21.22288797 \\
10 & Kottayam        & 697.3226977 \\
 4 & Kozhikode       & 68.53598093 \\
 5 & Malappuram      & 187.1457862 \\
 6 & Palakkad        & 86.78447476 \\
12 & Pathanamthitta  & 91.15051287 \\
 7 & Thrissur        & 464.847006 \\
 3 & Wayanad         & 24.260779 \\
\bottomrule
\end{tabular}
\begin{tablenotes}[para,flushleft]
\footnotesize
Notes: “Exposure” is reported in same unit as the derived sub-district level flood is in hectares.
\end{tablenotes}
\end{threeparttable}
\end{table}

\section{Descriptive Statistics}\label{app:desc-stat}

Table~\ref{tab:sumstats_cohensd_pretreatment} presents pre-flood period summary statistics for continuous variables across districts classified as high exposure, low exposure, and control groups, separately for men and women. The key takeaway is that the standardized mean differences between treatment and control groups are small (except women's years of education). This implies that there is strong baseline comparability between treatment and control groups. Hence, Tamil Nadu is valid counterfactual.

Within all three groups of districts, men's paid labor outcomes are much higher compared to women. Extensive margin is captured by employment participation and intensive margin is measured by working hours. Both men and women exhibit lowest participation within high exposed districts (56\% men and 17\% women) and highest within control group which were never exposed to flood (63\% men and 22\% women). Similar trends are observed in terms of working hours.

Demographic characteristics are broadly similar across groups except women's years of education. Average age is about 37–38 years for men and women in Kerala, compared to 35–36 years in Tamil Nadu. Mean years of formal education vary between 8.3-8.7 years for men and women in Kerala, while in Tamil Nadu an average man spends 8.5 years in education while women spends 7.3 years. It should be noted that Kerala is the only state with highest educational attainment and 100\% literacy in India.

\begin{table}[h]\centering
\caption{Summary Statistics of Continuous Variables (Pre-Treatment)}
\label{tab:sumstats_cohensd_pretreatment}
\begin{threeparttable}
\resizebox{\textwidth}{!}{%
\begin{tabular}{lcccccccccc}
\toprule
& \multicolumn{2}{c}{High Exposure} & \multicolumn{2}{c}{Low Exposure} & \multicolumn{2}{c}{Control (TN)} & \multicolumn{2}{c}{SMD vs Control (High)} & \multicolumn{2}{c}{SMD vs Control (Low)} \\
\cmidrule(lr){2-3}\cmidrule(lr){4-5}\cmidrule(lr){6-7}\cmidrule(lr){8-9}\cmidrule(lr){10-11}
& Men & Women & Men & Women & Men & Women & Men & Women & Men & Women \\
\midrule
Paid labor (share) &
0.56 & 0.17 & 0.56 & 0.21 & 0.63 & 0.22 & -0.14 & -0.15 & -0.13 & -0.04 \\
& (0.50) & (0.37) & (0.50) & (0.41) & (0.48) & (0.42) & & & & \\[0.5em]

Total hours worked &
25.20 & 6.10 & 24.61 & 6.14 & 29.85 & 8.91 & -0.17 & -0.15 & -0.19 & -0.15 \\
& (27.03) & (16.82) & (26.87) & (16.33) & (28.32) & (18.98) & & & & \\[0.5em]

Age &
37.63 & 38.53 & 36.67 & 38.44 & 35.08 & 35.69 & 0.13 & 0.14 & 0.08 & 0.14 \\
& (20.87) & (20.79) & (20.50) & (20.03) & (18.80) & (18.50) & & & & \\[0.5em]

Years of education &
8.69 & 8.65 & 8.40 & 8.28 & 8.47 & 7.25 & 0.05 & 0.27 & -0.02 & 0.20 \\
& (4.21) & (4.57) & (4.13) & (4.63) & (4.90) & (5.26) & & & & \\
\bottomrule
\end{tabular}
}
\begin{tablenotes}[para,flushleft]
\footnotesize
Notes: Means are reported with standard deviations in parentheses. Data are from the pre-treatment period.\\ High exposure group includes those districts with $\text{Exposure} \geq 91.15$. Low exposure group covers those with\\$\text{Exposure} < 91.15$. Tamil Nadu is the control group. Standardized mean differences (SMDs) are calculated\\ as Cohen's d using the pooled standard deviation.
\end{tablenotes}
\end{threeparttable}
\end{table}

\newpage 

\section{Detailed Results: Table}
Details about the estimated coefficients are presented from the next page onwards.
\begin{table}\centering
\scriptsize
\ignorespacesafterend
\caption{Event Study Estimates in Highly Exposed Districts as Treatment Group}
\label{tab:eventstudy_combined}
\begin{threeparttable}
\setlength{\tabcolsep}{6pt}
\begin{tabular}{lcccc}
\toprule
& \multicolumn{2}{c}{Men} & \multicolumn{2}{c}{Women} \\
\cmidrule(lr){2-3}\cmidrule(lr){4-5}
& (1)  & (2)  & (3)  & (4)  \\
\midrule
\multicolumn{5}{l}{\textit{Panel A: Paid employment (extensive margin)}} \\
\midrule
Lead 4 & 0.014 & 0.007 & -0.037 & -0.037 \\
       & (0.036) & (0.028) & (0.025) & (0.025) \\
Lead 3 & -0.005 & -0.018 & -0.008 & 0.003 \\
       & (0.040) & (0.029) & (0.029) & (0.028) \\
Lead 2 & -0.032 & -0.022 & -0.041 & -0.027 \\
       & (0.037) & (0.027) & (0.026) & (0.025) \\
Lag 0  & -0.060\sym{*} & -0.041\sym{*} & -0.015 & -0.001 \\
       & (0.033) & (0.022) & (0.023) & (0.023) \\
Lag 1  & 0.016 & 0.042\sym{*} & -0.040\sym{*} & -0.028 \\
       & (0.033) & (0.023) & (0.024) & (0.024) \\
Lag 2  & 0.015 & 0.023 & 0.033 & -0.029 \\
       & (0.034) & (0.024) & (0.025) & (0.024) \\
Lag 3  & 0.056\sym{*} & 0.038 & -0.007 & -0.020 \\
       & (0.032) & (0.023) & (0.024) & (0.024) \\
\midrule
Observations & 34,925 & 34,925 & 36,605 & 36,605 \\
$R^{2}$ & 0.027 & 0.346 & 0.028 & 0.083 \\
Household-level Controls    & \tikzxmark & \checkmark & \tikzxmark & \checkmark \\
Individual-level Controls   & \tikzxmark & \checkmark & \tikzxmark & \checkmark \\
District Fixed Effects & \checkmark & \checkmark & \checkmark & \checkmark \\
Linear time trend & \checkmark & \checkmark & \checkmark & \checkmark \\
\midrule
\multicolumn{5}{l}{\textit{Panel B: Weekly hours of paid work (intensive margin)}} \\
\midrule
Lead 4 & -0.867 & -0.861 & -1.833 & -1.807 \\
       & (1.932) & (1.553) & (1.141) & (1.123) \\
Lead 3 & -0.788 & -1.363 & -0.643 & -0.177 \\
       & (2.216) & (1.775) & (1.298) & (1.262) \\
Lead 2 & -0.406 & 0.118 & -0.945 & -0.304 \\
       & (2.032) & (1.583) & (1.221) & (1.197) \\
Lag 0  & -4.729\sym{***} & -3.748\sym{***} & -1.076 & -0.513 \\
       & (1.709) & (1.252) & (1.063) & (1.050) \\
Lag 1  & -0.822 & 0.408 & -1.411 & -0.826 \\
       & (1.769) & (1.301) & (1.072) & (1.051) \\
Lag 2  & -2.864 & -2.298\sym{*} & -2.397\sym{**} & -2.197\sym{**} \\
       & (1.782) & (1.341) & (1.071) & (1.052) \\
Lag 3  & -1.234 & -1.793 & -1.587 & -1.927\sym{*} \\
       & (1.714) & (1.300) & (1.028) & (1.017) \\
\midrule
Observations & 34,468 & 34,468 & 35,458 & 35,458 \\
$R^{2}$ & 0.029 & 0.346 & 0.089 & 0.076 \\
Household-level Controls    & \tikzxmark & \checkmark & \tikzxmark & \checkmark \\
Individual-level Controls   & \tikzxmark & \checkmark & \tikzxmark & \checkmark \\
District Fixed Effects & \checkmark & \checkmark & \checkmark & \checkmark \\
Linear time trend & \checkmark & \checkmark & \checkmark & \checkmark \\
\bottomrule
\end{tabular}
\begin{tablenotes}[para,flushleft]
\footnotesize
Notes: Robust standard errors in parentheses. \sym{*} $p<0.10$, \sym{**} $p<0.05$, \sym{***} $p<0.01$. 
Columns (1) and (3) include only district fixed effects, while columns (2) and (4) add individual-level controls.
\end{tablenotes}
\end{threeparttable}
\end{table}

\begin{table}[!htbp]\centering
\scriptsize
\caption{Event Study Estimates in Less Exposed Districts as Treatment Group}
\label{tab:lessaffected_combined}
\begin{threeparttable}
\setlength{\tabcolsep}{6pt}
\begin{tabular}{lcccc}
\toprule
& \multicolumn{2}{c}{Men} & \multicolumn{2}{c}{Women} \\
\cmidrule(lr){2-3}\cmidrule(lr){4-5}
& (1) & (2) & (3)  & (4) \\
\midrule
\multicolumn{5}{l}{\textit{Panel A: Paid employment (extensive margin)}} \\
\midrule
Lead 4 & -0.075\sym{**} & -0.049\sym{*} & -0.051\sym{**} & -0.038 \\
       & (0.033)        & (0.027)       & (0.025)        & (0.024) \\
Lead 3 & -0.026         & -0.041        & -0.022         & -0.017 \\
       & (0.032)        & (0.025)       & (0.024)        & (0.023) \\
Lead 2 & -0.026         & -0.029        & 0.013          & 0.010 \\
       & (0.032)        & (0.026)       & (0.025)        & (0.024) \\
Lag 0  & 0.003          & -0.009        & 0.016          & 0.019 \\
       & (0.028)        & (0.023)       & (0.022)        & (0.022) \\
Lag 1  & 0.049\sym{*}   & 0.024         & -0.026         & -0.035\sym{*} \\
       & (0.027)        & (0.022)       & (0.021)        & (0.021) \\
Lag 2  & 0.008          & 0.002         & -0.017         & -0.020 \\
       & (0.028)        & (0.023)       & (0.022)        & (0.021) \\
Lag 3  & -0.018         & -0.019        & -0.038\sym{*}  & -0.035 \\
       & (0.029)        & (0.023)       & (0.022)        & (0.021) \\
\midrule
Observations & 36{,}797 & 36{,}797 & 38{,}799 & 38{,}799 \\
$R^{2}$      & 0.008 & 0.338 & 0.023 & 0.080 \\
Household-level Controls    & \tikzxmark & \checkmark & \tikzxmark & \checkmark \\
Individual-level Controls   & \tikzxmark & \checkmark & \tikzxmark & \checkmark \\
District Fixed Effects      & \checkmark & \checkmark & \checkmark & \checkmark \\
Linear time trend           & \checkmark & \checkmark & \checkmark & \checkmark \\
\midrule
\multicolumn{5}{l}{\textit{Panel B: Weekly working hours (intensive margin)}} \\
\midrule
Lead 4 & -2.898\sym{*} & -1.749 & -0.119 & 0.330 \\
       & (1.678)       & (1.394) & (1.006) & (0.970) \\
Lead 3 & 1.560         & 0.601   & 1.156   & 1.329 \\
       & (1.694)       & (1.401) & (0.995) & (0.966) \\
Lead 2 & 0.199         & 0.120   & 0.778   & 0.616 \\
       & (1.650)       & (1.377) & (1.016) & (0.981) \\
Lag 0  & -2.347\sym{*} & -2.875\sym{**} & 0.095 & 0.083 \\
       & (1.384)       & (1.201)        & (0.853) & (0.838) \\
Lag 1  & 0.566         & -0.391  & -0.544  & -0.913 \\
       & (1.349)       & (1.193) & (0.812) & (0.796) \\
Lag 2  & -0.428        & -0.486  & -0.233  & -0.295 \\
       & (1.397)       & (1.184) & (0.824) & (0.799) \\
Lag 3  & -0.861        & -0.742  & -1.230  & -1.118 \\
       & (1.451)       & (1.203) & (0.845) & (0.823) \\
\midrule
Observations & 36{,}334 & 36{,}334 & 37{,}702 & 37{,}702 \\
$R^{2}$      & 0.029 & 0.344 & 0.029 & 0.076 \\
Household-level Controls    & \tikzxmark & \checkmark & \tikzxmark & \checkmark \\
Individual-level Controls   & \tikzxmark & \checkmark & \tikzxmark & \checkmark \\
District Fixed Effects      & \checkmark & \checkmark & \checkmark & \checkmark \\
Linear time trend           & \checkmark & \checkmark & \checkmark & \checkmark \\
\bottomrule
\end{tabular}
\begin{tablenotes}[para,flushleft]
\footnotesize
Notes: Robust standard errors in parentheses. \sym{*} $p<0.10$, \sym{**} $p<0.05$, \sym{***} $p<0.01$. 
Columns (1) and (3) include only district fixed effects, while columns (2) and (4) add individual-level controls.
\end{tablenotes}
\end{threeparttable}
\end{table}

\begin{table}[!htbp]\centering
\scriptsize
\caption{Event Study Estimates with 75th Percentile Threshold as Treatment}
\label{tab:robust_p75_full}
\begin{threeparttable}
\setlength{\tabcolsep}{6pt}
\begin{tabular}{lcccc}
\toprule
& \multicolumn{2}{c}{Men} & \multicolumn{2}{c}{Women} \\
\cmidrule(lr){2-3}\cmidrule(lr){4-5}
& (1) & (2)  & (3)  & (4)  \\
\midrule
\multicolumn{5}{l}{\textit{Panel A: Paid employment (extensive margin)}} \\
\midrule
Lead 4 &  \phantom{-}0.007 & -0.002 & -0.033 & -0.033 \\
       & (0.037)           & (0.027) & (0.026) & (0.025) \\
Lead 3 & -0.015 & -0.030 & -0.017 & -0.009 \\
       & (0.041) & (0.030) & (0.029) & (0.028) \\
Lead 2 & -0.038 & -0.031 & -0.030 & -0.015 \\
       & (0.037) & (0.027) & (0.027) & (0.026) \\
Lag 0  & -0.082\sym{**} & -0.054\sym{**} & -0.010 & \phantom{-}0.007 \\
       & (0.033)         & (0.022)        & (0.025) & (0.024) \\
Lag 1  & -0.006 & \phantom{-}0.017 & -0.047\sym{*} & -0.026 \\
       & (0.034) & (0.023) & (0.025) & (0.024) \\
Lag 2  & \phantom{-}0.005 & -0.003 & -0.020 & -0.021 \\
       & (0.034)          & (0.024) & (0.026) & (0.025) \\
Lag 3  & \phantom{-}0.054 & \phantom{-}0.034 & -0.013 & -0.024 \\
       & (0.033)          & (0.023) & (0.024) & (0.024) \\
\midrule
Observations & 34{,}173 & 34{,}173 & 35{,}690 & 35{,}690 \\
$R^{2}$      & 0.010 & 0.345 & 0.028 & 0.084 \\
District FE           & \checkmark & \checkmark & \checkmark & \checkmark \\
Linear time trend     & \checkmark & \checkmark & \checkmark & \checkmark \\
Individual-level ctrls& \tikzxmark & \checkmark & \tikzxmark & \checkmark \\
\midrule
\multicolumn{5}{l}{\textit{Panel B: Weekly working hours (intensive margin)}} \\
\midrule
Lead 4 & -0.855 & -1.016 & -1.455 & -1.418 \\
       & (1.954) & (1.554) & (1.143) & (1.120) \\
Lead 3 & -0.976 & -1.627 & \phantom{-}0.092 & \phantom{-}0.435 \\
       & (2.270) & (1.812) & (1.344) & (1.302) \\
Lead 2 & -0.314 & \phantom{-}0.044 & -0.314 & \phantom{-}0.290 \\
       & (2.049) & (1.620) & (1.245) & (1.214) \\
Lag 0  & -5.509\sym{***} & -4.113\sym{***} & -0.685 & -0.007 \\
       & (1.736)          & (1.254)         & (1.085) & (1.064) \\
Lag 1  & -0.891 & -0.053 & -1.379 & -0.551 \\
       & (1.815) & (1.322) & (1.071) & (1.045) \\
Lag 2  & -3.420\sym{*} & -3.579\sym{**} & -1.909\sym{*} & -1.904\sym{*} \\
       & (1.823)        & (1.373)        & (1.087)        & (1.061) \\
Lag 3  & -1.007 & -1.630 & -1.431 & -1.693\sym{*} \\
       & (1.728) & (1.290) & (1.019) & (1.001) \\
\midrule
Observations & 33{,}718 & 33{,}718 & 34{,}564 & 34{,}564 \\
$R^{2}$      & 0.030 & 0.347 & 0.028 & 0.076 \\
District FE           & \checkmark & \checkmark & \checkmark & \checkmark \\
Linear time trend     & \checkmark & \checkmark & \checkmark & \checkmark \\
Individual-level ctrls& \tikzxmark & \checkmark & \tikzxmark & \checkmark \\
\bottomrule
\end{tabular}
\begin{tablenotes}[para,flushleft]
\footnotesize
Notes: Robust standard errors in parentheses. \sym{*} $p<0.10$, \sym{**} $p<0.05$, \sym{***} $p<0.01$. Treatment equals 1 for districts at or above the 75th percentile of the exposure distribution; controls are all other districts. Columns (2) and (4) include individual-level controls; all columns include district fixed effects and a linear time trend.
\end{tablenotes}
\end{threeparttable}
\end{table}

\begin{table}[!htbp]\centering
\scriptsize
\caption{Event Study Estimates by Dropping Border Districts}
\label{tab:spillover_robust}
\begin{threeparttable}
\setlength{\tabcolsep}{6pt}
\begin{tabular}{lcccc}
\toprule
& \multicolumn{2}{c}{Men} & \multicolumn{2}{c}{Women} \\
\cmidrule(lr){2-3}\cmidrule(lr){4-5}
& (1) & (2)  & (3)  & (4)  \\
\midrule
\multicolumn{5}{l}{\textit{Panel A: Paid employment (extensive margin)}} \\
\midrule
Lead 4 & -0.008 & -0.012 & -0.047\sym{**} & -0.048\sym{**} \\
       & (0.029) & (0.022) & (0.020) & (0.020) \\
Lead 3 & -0.041 & -0.058\sym{**} & -0.058\sym{***} & -0.048\sym{**} \\
       & (0.030) & (0.022) & (0.021) & (0.020) \\
Lead 2 & -0.038 & -0.034 & -0.012 & -0.005 \\
       & (0.028) & (0.022) & (0.021) & (0.020) \\
Lag 0  & -0.059\sym{**} & -0.048\sym{**} & 0.007 & 0.011 \\
       & (0.026) & (0.018) & (0.020) & (0.019) \\
Lag 1  & 0.046\sym{*} & 0.044\sym{**} & -0.028 & -0.025 \\
       & (0.025) & (0.018) & (0.019) & (0.019) \\
Lag 2  & 0.013 & -0.001 & -0.005 & -0.011 \\
       & (0.026) & (0.019) & (0.020) & (0.019) \\
Lag 3  & 0.050\sym{**} & 0.023 & -0.025 & -0.036\sym{*} \\
       & (0.025) & (0.019) & (0.019) & (0.019) \\
\midrule
Observations & 38{,}287 & 38{,}287 & 40{,}356 & 40{,}356 \\
$R^{2}$      & 0.010    & 0.347    & 0.027    & 0.083 \\
District Fixed Effects    & \checkmark & \checkmark & \checkmark & \checkmark \\
Linear time trend         & \checkmark & \checkmark & \checkmark & \checkmark \\
Individual-level controls & \tikzxmark & \checkmark & \tikzxmark & \checkmark \\
\midrule
\multicolumn{5}{l}{\textit{Panel B: Weekly working hours (intensive margin)}} \\
\midrule
Lead 4 & -1.049 & -1.034 & -0.974 & -1.008 \\
       & (1.462) & (1.170) & (0.870) & (0.849) \\
Lead 3 & -0.416 & -1.403 & -0.729 & -0.365 \\
       & (1.635) & (1.284) & (0.907) & (0.884) \\
Lead 2 & 0.148  & 0.308  & 0.476  & 0.765 \\
       & (1.507) & (1.217) & (0.941) & (0.913) \\
Lag 0  & -3.247\sym{**} & -2.616\sym{**} & 0.272 & 0.392 \\
       & (1.296)        & (0.971)        & (0.831) & (0.817) \\
Lag 1  & 1.265 & 1.256 & -0.633 & -0.546 \\
       & (1.305) & (1.015) & (0.780) & (0.762) \\
Lag 2  & -0.872 & -1.329 & -0.461 & -0.669 \\
       & (1.322) & (1.020) & (0.797) & (0.773) \\
Lag 3  & 0.960 & -0.015 & -1.397\sym{*} & -1.725\sym{**} \\
       & (1.311) & (1.006) & (0.767) & (0.752) \\
\midrule
Observations & 37{,}816 & 37{,}816 & 39{,}196 & 39{,}196 \\
$R^{2}$      & 0.029    & 0.345    & 0.029    & 0.076 \\
District Fixed Effects    & \checkmark & \checkmark & \checkmark & \checkmark \\
Linear time trend         & \checkmark & \checkmark & \checkmark & \checkmark \\
Individual-level controls & \tikzxmark & \checkmark & \tikzxmark & \checkmark \\
\bottomrule
\end{tabular}
\begin{tablenotes}[para,flushleft]
\footnotesize
Notes: Robust standard errors in parentheses. \sym{*} $p<0.10$, \sym{**} $p<0.05$, \sym{***} $p<0.01$. 
All specifications drop border districts to assess spatial spillovers. Columns (1) and (3) omit individual-level controls; columns (2) and (4) include them. All columns include district fixed effects and a linear time trend.
\end{tablenotes}
\end{threeparttable}
\end{table}

\begin{table}[!htbp]\centering
\scriptsize
\caption{Event Study Estimates by Marital Status}
\label{tab:eventstudy_marital}
\begin{threeparttable}
\setlength{\tabcolsep}{6pt}
\begin{tabular}{lcccc}
\toprule
& \multicolumn{2}{c}{Men} & \multicolumn{2}{c}{Women} \\
\cmidrule(lr){2-3}\cmidrule(lr){4-5}
& Unmarried & Married & Unmarried & Married \\
& (1) & (2) & (3) & (4) \\
\midrule
\multicolumn{5}{l}{\textit{Panel A: Paid employment (extensive margin)}} \\
\midrule
Lead 4 & 0.053 & -0.042 & 0.002 & -0.041 \\
       & (0.045) & (0.034) & (0.032) & (0.032) \\
Lead 3 & 0.014 & -0.031 & -0.014 & 0.019 \\
       & (0.048) & (0.034) & (0.035) & (0.037) \\
Lead 2 & 0.016 & -0.051 & -0.052\sym{*} & -0.022 \\
       & (0.045) & (0.033) & (0.031) & (0.034) \\
Lag 0  & -0.041 & -0.022 & 0.003 & 0.004 \\
       & (0.038) & (0.028) & (0.030) & (0.030) \\
Lag 1  & 0.030 & 0.079\sym{***} & 0.014 & -0.030 \\
       & (0.038) & (0.027) & (0.031) & (0.031) \\
Lag 2  & 0.054 & 0.013 & 0.020 & -0.030 \\
       & (0.038) & (0.028) & (0.028) & (0.031) \\
Lag 3  & 0.062 & 0.026 & 0.030 & -0.028 \\
       & (0.040) & (0.027) & (0.029) & (0.031) \\
\midrule
Observations & 15,152 & 19,272 & 11,333 & 25,272 \\
$R^{2}$ & 0.43 & 0.24 & 0.23 & 0.08 \\
District FE & \checkmark & \checkmark & \checkmark & \checkmark \\
Linear trend & \checkmark & \checkmark & \checkmark & \checkmark \\
\midrule
\multicolumn{5}{l}{\textit{Panel B: Weekly hours of paid work (intensive margin)}} \\
\midrule
Lead 4 & 1.828 & -3.229 & -0.731 & -1.870 \\
       & (2.236) & (2.064) & (1.588) & (1.424) \\
Lead 3 & 0.424 & -1.820 & -0.775 & 0.379 \\
       & (2.281) & (2.493) & (1.564) & (1.654) \\
Lead 2 & -0.098 & 0.152 & -2.101 & 0.163 \\
       & (2.189) & (2.221) & (1.551) & (1.608) \\
Lag 0  & -2.859\sym{*} & -3.485\sym{*} & 0.592 & -0.846 \\
       & (1.726) & (1.842) & (1.552) & (1.339) \\
Lag 1  & 0.461 & 1.478 & -0.658 & -0.308 \\
       & (1.738) & (1.895) & (1.422) & (1.365) \\
Lag 2  & -1.011 & -2.395 & -0.093 & -2.360\sym{*} \\
       & (1.742) & (1.900) & (1.361) & (1.356) \\
Lag 3  & 0.849 & -3.066\sym{*} & 0.450 & -2.488\sym{*} \\
       & (1.801) & (1.819) & (1.376) & (1.308) \\
\midrule
Observations & 14,910 & 19,558 & 11,317 & 24,141 \\
$R^{2}$ & 0.32 & 0.22 & 0.17 & 0.06 \\
District FE & \checkmark & \checkmark & \checkmark & \checkmark \\
Linear trend & \checkmark & \checkmark & \checkmark & \checkmark \\
\bottomrule
\end{tabular}
\begin{tablenotes}[para,flushleft]
\footnotesize
Notes: Robust standard errors in parentheses. 
\sym{*} $p<0.10$, \sym{**} $p<0.05$, \sym{***} $p<0.01$. 
All regressions include district fixed effects and a linear time trend. 
\end{tablenotes}
\end{threeparttable}
\end{table}

\newpage
\begin{table}[!htbp]\centering
\scriptsize
\caption{Event Study Estimates by Household Dependency}
\label{tab:eventstudy_depburden}
\begin{threeparttable}
\setlength{\tabcolsep}{6pt}
\begin{tabular}{lcccc}
\toprule
& \multicolumn{2}{c}{Men} & \multicolumn{2}{c}{Women} \\
\cmidrule(lr){2-3}\cmidrule(lr){4-5}
& Low dep. & High dep. & Low dep. & High dep. \\
& (1) & (2) & (3) & (4) \\
\midrule
\multicolumn{5}{l}{\textit{Panel A: Paid employment (extensive margin)}} \\
\midrule
Lead 4 & -0.002 & 0.029 & -0.004 & -0.082\sym{**} \\
       & (0.039) & (0.039) & (0.038) & (0.037) \\
Lead 3 & -0.002 & 0.018 & 0.022 & -0.037 \\
       & (0.040) & (0.045) & (0.041) & (0.043) \\
Lead 2 & -0.037 & 0.016 & 0.006 & -0.073\sym{*} \\
       & (0.039) & (0.041) & (0.037) & (0.041) \\
Lag 0  & -0.083\sym{***} & 0.015 & 0.023 & -0.032 \\
       & (0.033) & (0.031) & (0.033) & (0.036) \\
Lag 1  & -0.005 & 0.107\sym{***} & 0.008 & -0.074\sym{**} \\
       & (0.033) & (0.033) & (0.035) & (0.037) \\
Lag 2  & 0.014 & 0.046 & 0.009 & -0.083\sym{**} \\
       & (0.034) & (0.036) & (0.035) & (0.038) \\
Lag 3  & 0.031 & 0.070\sym{**} & 0.010 & -0.052 \\
       & (0.033) & (0.033) & (0.034) & (0.037) \\
\midrule
Observations & 21{,}374 & 11{,}024 & 21{,}375 & 12{,}562 \\
$R^{2}$      & 0.22 & 0.55 & 0.06 & 0.14 \\
District Fixed Effects & \checkmark & \checkmark & \checkmark & \checkmark \\
Linear time trend      & \checkmark & \checkmark & \checkmark & \checkmark \\
\midrule
\multicolumn{5}{l}{\textit{Panel B: Weekly hours of paid work (intensive margin)}} \\
\midrule
Lead 4 & -1.206 & 0.010 & -1.742 & -2.573 \\
       & (2.276) & (2.075) & (1.77) & (1.599) \\
Lead 3 & -1.242 & 0.361 & 0.115 & -1.734 \\
       & (2.449) & (2.845) & (1.908) & (1.721) \\
Lead 2 & -1.901 & 3.056 & 0.178 & -2.246 \\
       & (2.333) & (2.358) & (1.802) & (1.827) \\
Lag 0  & -5.949\sym{***} & -1.521 & -0.883 & -0.257 \\
       & (1.837) & (1.762) & (1.594) & (1.597) \\
Lag 1  & -1.834 & 2.647 & -0.985 & -1.419 \\
       & (1.927) & (1.835) & (1.596) & (1.610) \\
Lag 2  & -3.398\sym{*} & -0.918 & -1.358 & -4.00\sym{**} \\
       & (1.967) & (1.861) & (1.623) & (1.532) \\
Lag 3  & -2.687 & -0.374 & -2.1 & -2.314 \\
       & (1.868) & (1.866) & (1.526) & (1.516) \\
\midrule
Observations & 21{,}031 & 10{,}936 & 20{,}580 & 12{,}290 \\
$R^{2}$      & 0.251 & 0.525 & 0.07 & 0.11 \\
District Fixed Effects & \checkmark & \checkmark & \checkmark & \checkmark \\
Linear time trend      & \checkmark & \checkmark & \checkmark & \checkmark \\
\bottomrule
\end{tabular}
\begin{tablenotes}[para,flushleft]
\footnotesize
Notes: Robust standard errors in parentheses. \sym{*}$p<0.10$, \sym{**}$p<0.05$, \sym{***}$p<0.01$. 
All specifications include district fixed effects and a linear time trend.
Men/High dep. and Women/High dep. columns correspond to households with higher dependency burden; Low dep. to households with lower dependency burden.
\end{tablenotes}
\end{threeparttable}
\end{table}
\newpage
\begin{table}[!htbp]\centering
\scriptsize
\caption{Event Study Estimates by Employment Sector}
\label{tab:eventstudy_bysector}
\begin{threeparttable}
\setlength{\tabcolsep}{5.5pt}
\begin{tabular}{lcccccc}
\toprule
& \multicolumn{3}{c}{Men} & \multicolumn{3}{c}{Women} \\
\cmidrule(lr){2-4}\cmidrule(lr){5-7}
& Primary & Secondary & Tertiary & Primary & Secondary & Tertiary \\
& (1) & (2) & (3) & (4) & (5) & (6) \\
\midrule
\multicolumn{7}{l}{\textit{Panel A: Paid employment (extensive margin)}} \\
\midrule
Lead 4 & -0.025 & -0.0064 & -0.001 & 0.294 & -0.031 & -0.061 \\
       & (0.022) & (0.0066) & (0.0059) & (0.194) & (0.088) & (0.0412) \\
Lead 3 & -0.022 & -0.0049 & 0.002 & -0.041 & 0.060 & 0.012 \\
       & (0.021) & (0.0039) & (0.0038) & (0.195) & (0.064) & (0.0188) \\
Lead 2 & 0.045 & -0.0096 & -0.0080 & 0.080 & 0.030 & -0.142\sym{**} \\
       & (0.034) & (0.0093) & (0.0088) & (0.226) & (0.069) & (0.0679) \\
Lag 0  & -0.001 & -0.0107\sym{**} & -0.016\sym{*} & 0.192 & 0.027 & 0.0019 \\
       & (0.014) & (0.0051) & (0.0094) & (0.182) & (0.056) & (0.0205) \\
Lag 1  & 0.008 & -0.0149\sym{*} & -0.0005 & 0.169 & -0.137 & -0.069\sym{**} \\
       & (0.017) & (0.0085) & (0.0033) & (0.180) & (0.094) & (0.0293) \\
Lag 2  & 0.002 & -0.0041 & -0.020\sym{**} & 0.181 & 0.075 & -0.0415\sym{*} \\
       & (0.021) & (0.0036) & (0.0098) & (0.172) & (0.055) & (0.0249) \\
Lag 3  & 0.0336\sym{**} & -0.0179\sym{**} & -0.015\sym{**} & 0.170 & 0.042 & -0.0937\sym{***} \\
       & (0.017) & (0.0079) & (0.0070) & (0.167) & (0.056) & (0.0347) \\
\midrule
Observations & 4{,}263 & 6{,}874 & 9{,}077 & 2{,}640 & 2{,}670 & 3{,}483 \\
$R^{2}$      & 0.23 & 0.03 & 0.04 & 0.15 & 0.13 & 0.12 \\
District Fixed Effects & \checkmark & \checkmark & \checkmark & \checkmark & \checkmark & \checkmark \\
Linear time trend      & \checkmark & \checkmark & \checkmark & \checkmark & \checkmark & \checkmark \\
\midrule
\multicolumn{7}{l}{\textit{Panel B: Weekly hours of paid work (intensive margin)}} \\
\midrule
Lead 4 & -2.799 & 0.479 & -4.755\sym{**} & 10.475\sym{*} & 8.632\sym{*} & 0.864 \\
       & (3.000) & (2.155) & (2.017) & (5.843) & (5.038) & (3.379) \\
Lead 3 & 1.538 & -4.446\sym{*} & 2.280 & 6.920 & 1.621 & -2.107 \\
       & (2.754) & (2.577) & (2.296) & (5.768) & (4.856) & (3.540) \\
Lead 2 & 4.633 & 1.065 & -0.316 & 9.893 & 4.353 & 3.545 \\
       & (3.057) & (2.460) & (1.963) & (6.556) & (4.336) & (3.562) \\
Lag 0  & -6.044\sym{**} & -7.201\sym{***} & -3.234\sym{*} & 8.228 & 0.136 & -6.427\sym{**} \\
       & (2.383) & (1.916) & (1.780) & (6.329) & (4.523) & (3.040) \\
Lag 1  & -6.705\sym{***} & -2.711 & -2.213 & 16.708\sym{***} & -2.452 & -1.587 \\
       & (2.362) & (2.081) & (1.886) & (4.298) & (4.048) & (2.874) \\
Lag 2  & 0.839 & -4.234\sym{**} & -4.905\sym{**} & 5.403\sym{*} & 4.235 & -7.199\sym{**} \\
       & (2.566) & (1.834) & (2.088) & (3.265) & (4.009) & (3.014) \\
Lag 3  & -6.587\sym{***} & -6.395\sym{***} & -4.850\sym{***} & 6.447 & -1.611 & -11.620\sym{***} \\
       & (2.270) & (1.776) & (1.832) & (4.322) & (4.038) & (3.225) \\
\midrule
Observations & 4{,}057 & 6{,}786 & 8{,}914 & 2{,}044 & 2{,}444 & 3{,}158 \\
$R^{2}$      & 0.18 & 0.23 & 0.09 & 0.23 & 0.18 & 0.18 \\
District Fixed Effects & \checkmark & \checkmark & \checkmark & \checkmark & \checkmark & \checkmark \\
Linear time trend      & \checkmark & \checkmark & \checkmark & \checkmark & \checkmark & \checkmark \\
\bottomrule
\end{tabular}
\begin{tablenotes}[para,flushleft]
\footnotesize
Notes: Robust standard errors in parentheses. \sym{*} $p<0.10$, \sym{**} $p<0.05$, \sym{***} $p<0.01$. 
All specifications include district fixed effects and a linear time trend.
Panel A outcomes are proportions (0--1); Panel B outcomes are hours.
\end{tablenotes}
\end{threeparttable}
\end{table}

\end{document}